\documentclass[prd,preprint,onecolumn,12pt]{revtex4}%
\usepackage{aas_macros}
\usepackage{amsfonts}
\usepackage{amsmath}
\usepackage{graphicx}
\usepackage{amssymb}
\usepackage{appendix}%
\providecommand{\U}[1]{\protect\rule{.1in}{.1in}}
\begin{document}
\title{Thermalization of the mildly relativistic plasma}
\author{A.G. Aksenov}
\affiliation{Institute for Theoretical and Experimental Physics, B. Cheremushkinskaya 25,
117218 Moscow, Russia and Institute for Computer-Aided Design, Russian Academy
of Sciences, Vtoraya Brestskaya 19/18, Moscow, 123056, Russia}
\author{R. Ruffini}
\author{G.V. Vereshchagin}
\affiliation{ICRANet p.le della Repubblica, 10, 65100 Pescara,\ Italy}
\affiliation{ICRA and University of Rome \textquotedblleft Sapienza\textquotedblright,
Physics Department, p.le A. Moro 5, 00185 Rome, Italy}

\begin{abstract}
In the recent Letter \cite{2007PhRvL..99l5003A} we considered the approach of
nonequilibrium pair plasma towards thermal equilibrium state adopting a
kinetic treatment and solving numerically the relativistic Boltzmann
equations. It was shown that plasma in the energy range 0.1-10 MeV first
reaches kinetic equilibrium, on a timescale $t_{\mathrm{k}}\lesssim10^{-14}$
sec, with detailed balance between binary interactions such as Compton, Bhabha
and M{\o }ller scattering, and pair production and annihilation. Later the
electron-positron-photon plasma approaches thermal equilibrium on a timescale
$t_{\mathrm{th}}\lesssim10^{-12}$ sec, with detailed balance for all direct
and inverse reactions. In the present paper we systematically present details
of the computational scheme used in \cite{2007PhRvL..99l5003A}, as well as
generalize our treatment, considering proton loading of the pair plasma. When
proton loading is large, protons thermalize first by proton-proton scattering,
and then with the electron-positron-photon plasma by proton-electron
scattering. In the opposite case of small proton loading proton-electron
scattering dominates over proton-proton one. Thus in all cases the plasma,
even with proton admixture, reaches thermal equilibrium configuration on a
timescale $t_{\mathrm{th}}\lesssim10^{-11}$ sec. We show that it is crucial to
account for not only binary but also triple direct and inverse interactions
between electrons, positrons, photons and protons. Several explicit examples
are given and the corresponding timescales\ for reaching kinetic and thermal
equilibria are determined.

\end{abstract}
\maketitle

\section{Introduction}

An electron-positron plasma is of interest in many fields of physics and
astrophysics. One of the crucial quantities in this analysis is the timescale
of the thermalization process. In the early universe
\cite{1972gcpa.book.....W},\cite{1990eaun.book.....K}%
,\cite{1995PhDT..........H},\cite{2008gcpa.book.....W} during the lepton era,
ultrarelativistic electron-positron pairs contribute to the matter contents of
the Universe. In gamma-ray bursts (GRBs) electron-positron pairs play
essential role in the dynamics of expansion \cite{1986ApJ...308L..47G}%
,\cite{1999PhR...314..575P},\cite{1999A&A...350..334R}. Indications exist on
the presence of the pair plasma also in active galactic nuclei
\cite{1998Natur.395..457W}, in the center of our Galaxy
\cite{2005MNRAS.357.1377C},\ around hypothetical quark stars
\cite{1998PhRvL..80..230U}. In the laboratory pair plasma is expected to
appear in the fields of ultra intense lasers \cite{2006PhRvL..96n0402B}, where
particle production may serve as a diagnostic tool for high-energy plasma
\cite{kuznetsova:014027}.

In many stationary astrophysical sources the pair plasma is thought to be in
thermodynamic equilibrium. A detailed study of the relevant processes
\cite{1971SvA....15...17B},\cite{1976PhRvA..13.1563W}%
,\cite{1982ApJ...253..842L},\cite{1982ApJ...254..755G}%
,\cite{1983MNRAS.204.1269S},\cite{1990MNRAS.245..453C}, radiatiation
mechanisms \cite{1981ApJ...251..713L}, possible equilibrium configurations
\cite{1982ApJ...253..842L},\cite{1982ApJ...258..335S}%
,\cite{1985MNRAS.212..523G} and spectra \cite{1984ApJ...283..842Z} in an
optically thin pair plasma has been carried out. Particular attention has been
given to collisional relaxation process \cite{1981PhFl...24..102G}%
,\cite{1983MNRAS.202..467S}, pair production and annihilation
\cite{1982ApJ...258..321S}, relativistic bremsstrahlung
\cite{1980ApJ...238.1026G},\cite{1985A&A...148..386H}, double Compton
scattering \cite{1981ApJ...244..392L},\cite{1984ApJ...285..275G}.

An equilibrium occurs if the sum of all reaction rates vanishes. For instance,
electron-positron pairs are in equilibrium when the net pair production
(annihilation) rate is zero. This can be achieved by variety of ways and the
corresponding condition can be represented as a system of algebraic equations
\cite{1984MNRAS.209..175S}. However, the main assumption made in all the above
mentioned works is that the plasma is assumed to obey relativistic quantum
statistics. The latter is shown to be possible, in principle, in the range of
temperatures up to 10 MeV \cite{1971SvA....15...17B}%
,\cite{1983MNRAS.202..467S}. Our main task is to prove that independently of a
wide set of initial conditions, thermal equilibruim forms for the phase space
distribution functions are recovered during the process of thermalization by
two body and three body direct and inverse particle-particle collisions.

At the same time, in some cases mentioned above the pair plasma can be
optically thick. Although moderately thick plasmas have been considered in the
literature \cite{1985MNRAS.212..523G}, only qualitative description
\cite{1971SvA....15...17B},\cite{1982ApJ...258..335S} is available for large
optical depths. Assumption of thermal equilibrium is often adopted for rapidly
evolving systems such as GRBs without explicit proof
\cite{1986ApJ...308L..47G},\cite{1999PhR...314..575P}%
,\cite{1999A&A...350..334R},\cite{2004ApJ...601...78I}. Then hydrodynamic
approximation is usually applied both for leptons and photons. However,
particles may not be in equilibrium initially. Moreover, they may not reach an
equilibrium in rapidly evolving systems such as the early Universe or
transient events, when the energy is released on a very short timescale.

In the literature there is no consensus on this point. Some authors considered
thermal equilibrium as the initial state prior to expansion
\cite{1986ApJ...308L..47G},\cite{1999A&A...350..334R}, while others did not
\cite{1978MNRAS.183..359C}. In fact, the detailed study of the pair plasma
equilibrium configurations, performed in \cite{1982ApJ...258..335S}, cannot
answer this question, because essentially nonequilibrium processes have to be considered.

Thus, observations provide motivation for theoretical analysis of physical
conditions taking place in nonequilibrium optically thick pair plasma. Notice
that there is substantial difference between the ion-electron plasma on the
one hand and electron-positron plasma on the other hand. Firstly, the former
is collisionless in the wide range of parameters \cite{1981els..book.....L},
while collisions are always essential in the latter. Secondly, when collisions
are important relevant interactions in the former case are Coulomb scattering
of particles which are usually described by the classical Rutherford
cross-section. In contrast, interactions in the pair plasma are described by
quantum cross-sections even if the plasma itself can be still treated as
classical one.

Our study reported in \cite{2007PhRvL..99l5003A},\cite{2008AIPC..966..191A} in
the case of pure pair plasma clarified the issue of initial state of the pair
plasma in GRBs sources. Our numerical calculations show that the pair plasma
on a timescale $t\lesssim10^{-12}$ sec reach thermal equilibrium prior to
expansion, due to intense binary and triple collisions. In this paper we
present details about the computational scheme adopted in
\cite{2007PhRvL..99l5003A} and turn to a more general case, the pair plasma
loaded with baryons. Occurence of the thermalization process and the
corresponding timescales are necessary for determining the dynamics of GRBs.
Thermalization timescales $t\lesssim10^{-12}$ sec are indeed necessary in
order to relate the observed properties of GRBs to the nature of the source,
see e.g. \cite{2007AIPC..910...55R}.

In the next Section we give qualitative description of the pair plasma,
introducing some relevant parameters. In Section 3 we discuss pure pair
plasma. In Section 4 pair plasma with proton loading is discussed. In Section
5 we describe the computational scheme used in our analysis. In Section 6 we
present results of numerical computations. Discussion and conclusions follow
in the last Section. In Appendix A relevant conservation laws are recalled. In
Appendix B conditions for kinetic and thermal equilibria are formulated, and
the scheme for determination of temperatures and chemical potentials out of
number and energy densities are given. Binary interactions in the pair plasma
such as Compton, M{\o }ller and Bhabha scatterings, as well as pair creation
and annihilation by two photons are discussed in Appendix C. In Appendix D
Compton and Coulomb scatterings with protons are considered. In Appendix E
three-body radiative variants of the reactions listed above are given. Cutoff
scheme for numerical evaluation of emission and absorption coefficients are
presented in Appendix F. In Appendix G mass scaling of the matrix elements for
Coulomb scattering between electrons, positrons and protons is discussed. In
Appendix H the definition of matrix elements and cross-sections adopted in the
paper are given.

\section{Qualitative description of the pair plasma}

First of all we specify the domain of parameters characterizing the pair
plasma considered in this paper. It is convenient to use dimensionless
parameters usually adopted for this purpose.

We consider mildly relativistic pair plasma, thus the average energy per
particle $\epsilon$ brackets the electron rest mass energy%
\begin{equation}
0.1\lesssim\frac{\epsilon}{mc^{2}}\lesssim10\mathrm{.} \label{avenergy}%
\end{equation}
The lower boundary is required for significant concentrations of pairs, while
the upper boundary is set to avoid substantial production of other particles
such as muons and neutrinos.

We define the plasma parameter $\mathfrak{g}=(n_{-}d^{3})^{-1}$, where
$d=\sqrt{\frac{k_{B}T_{-}}{4\pi e^{2}n_{-}}}=\frac{c}{\omega}\sqrt{\theta_{-}}
$ is the Debye length, $k_{B}$ is Boltzmann's constant, $e$, $n_{-}$ and
$T_{-}$ are the electron charge, number density and temperature respectively,
$c$ the is speed of light, $\theta_{-}=k_{B}T_{-}/(mc^{2})$ is dimensionless
temperature, $\omega=\sqrt{4\pi e^{2}n_{-}/m}$ is the plasma frequency and $m$
is the electron mass. To ensure applicability of kinetic approach it is
necessary that the plasma parameter is small, $\mathfrak{g}\ll1$. This
condition means that kinetic energy of particles dominates their potential
energy due to mutual interaction. For the pair plasma considered in this paper
this condition is satisfied.

Further, the classicality parameter, defined as $\varkappa=e^{2}/(\hbar
v_{r})=\alpha/\beta_{r}$, where $\hbar$ is Planck's constant, $\alpha
=e^{2}/(\hbar c)$ is the fine structure constant, $v_{r}=\beta_{r}c$ is mean
relative velocity of particles, see (\ref{gammarel}) in Appendix. The
condition $\varkappa\gg1$ means that particles collisions can be considered
classically, while for $\varkappa\ll1$ quantum description is required. In our
case both for pairs and protons quantum cross-sections are used since
$\varkappa<1$.

The strength of screening of the Coulomb interactions is characterized by the
Coulomb logarithm $\Lambda=\mathcal{M}dv_{r}/\hbar$, where $\mathcal{M} $ is
the reduced mass. For electron-electron or electron-positron scattering the
reduced mass is just $m/2$, while for electron-proton or positron-proton
scattering the reduced mass is just the proton mass $\mathcal{M}\simeq M$; for
proton-proton scattering $\mathcal{M}\simeq M/2$. Coulomb logarithm varies
with mean particle velocity and Debye length, and it cannot be set a constant
as is usually done in most of studies of the pair plasma.

Finally, we consider pair plasma with linear dimensions $R$ exceeding the mean
free path of photons $l=\left(  n_{-}\sigma\right)  ^{-1}$, where $\sigma$ is
the corresponding total cross-section. Thus the optical depth $\tau=n\sigma
R\gg1$ is large, and interactions between photons and other particles have to
be taken in due account. We discuss these interaction in the next Section.

Note that natural parameters for perturbative expansion in the problem under
consideration are the fine structure constant $\alpha$ and the electron-proton
mass ratio $m/M$.

\section{Pure pair plasma}

For simplicity we first consider pure pair plasma composed of electrons
$e^{-}$, positrons $e^{+}$, and photons $\gamma$. We will turn to a more
general case, including protons $p$ in the next Section. We assume that pairs
or photons appear by some physical process in the region with a size $R $ and
on a timescale $t<R/c$. We further assume that distribution functions of
particles depend neither on spatial coordinates nor on the direction of
momenta. We then have $f_{i}=f_{i}(\epsilon,t)$, namely we consider isotropic
distributions functions in momentum space for a spatially uniform and
isotropic plasma.

To make sure that classical kinetic description is adequate we estimate the
dimensionless degeneracy temperature%
\begin{equation}
\theta_{F}=\left[  \left(  \frac{\hbar}{mc}\right)  ^{2}\left(  3\pi^{2}%
n_{-}\right)  ^{\frac{2}{3}}+1\right]  ^{1/2}-1,
\end{equation}
and compare it with the estimated temperature in thermal equilibrium. With our
initial conditions (\ref{avenergy}) the degeneracy temperature is always
smaller than the temperature in thermal equilibrium and therefore we can
safely apply the classical kinetic approach. Besides, since we deal with ideal
plasma with the plasma parameter $\mathfrak{g}\sim10^{-3}$ it is enough to
consider only one-particle distribution functions. These considerations
justify our computational approach based on classical relativistic Boltzmann
equation. At the same time the right hand side of Boltzmann equations contains
collisional integrals as functions of quantum matrix elements, as discussed
below and in Appendices C-E.

Relativistic Boltzmann equations \cite{1956DAN...107...807B}%
,\cite{1984oup..book.....M} in spherically symmetric case for which the
original code is designed \cite{2004ApJ...609..363A}\ are%
\begin{gather}
\frac{1}{c}\frac{\partial f_{i}}{\partial t}+\beta_{i}\left(  \mu
\frac{\partial f_{i}}{\partial r}+\frac{1-\mu^{2}}{r}\frac{\partial f_{i}%
}{\partial\mu}\right)  -\mathbf{\nabla}U\frac{\partial f_{i}}{\partial
\mathbf{p}}=\label{bol}\\
=\sum_{q}\left(  \eta_{i}^{q}-\chi_{i}^{q}f_{i}\right)  ,\nonumber
\end{gather}
where $\mu=\cos\vartheta$, $\vartheta$ is the angle between the radius vector
$\mathbf{r}$ from the origin and the particle momentum $\mathbf{p}$, $U$ is a
potential due to an external force, $\beta_{i}=v_{i}/c$ are particles
velocities, $f_{i}(\epsilon,t)$ are their distribution functions, the index
$i$ denotes the type of particle, $\epsilon$ is its energy, and $\eta_{i}^{q}$
and $\chi_{i}^{q}$ are the emission and the absorption coefficients for the
production of a particle of type \textquotedblleft$i$" via the physical
process labeled by $q$. This is a coupled system of
partial-integro-differential equations. For homogeneous and isotropic
distribution functions of electrons, positrons and photons (\ref{bol}) reduces
to%
\begin{equation}
\frac{1}{c}\frac{\partial f_{i}}{\partial t}=\sum_{q}\left(  \eta_{i}^{q}%
-\chi_{i}^{q}f_{i}\right)  , \label{BE}%
\end{equation}
which is a coupled system of integro-differential equations. In (\ref{BE}) we
also explicitly neglected the Vlasov term, describing collisionless
interaction of particles in the mean field, since energy density of
fluctuations of the electromagnetic field are many orders of magnitude smaller
than the energy density of particles \cite{PhysRevD.51.2677}.

Therefore, the left-hand side of the Boltzmann equation is reduced to partial
derivative of the distribution function with respect to time. The right-hand
side contains collisional integrals, representing interactions between
electrons, positrons and photons.

As example of collisional integral consider absorption coefficient for Compton
scattering which is given by%

\begin{equation}
\chi^{^{\mathrm{cs}}}f_{\gamma}=\int d\mathbf{k}^{\prime}d\mathbf{p}%
d\mathbf{p}^{\prime}W_{\mathbf{k}^{\prime},\mathbf{p}^{\prime};\mathbf{k}%
,\mathbf{p}}f_{\gamma}(\mathbf{k},t)f_{\pm}(\mathbf{p},t),
\label{comptonemission}%
\end{equation}
where $\mathbf{p}$ and $\mathbf{k}$ are momenta of electron (positron) and
photon respectively, $d\mathbf{p}=d\epsilon_{\pm}do\epsilon_{\pm}^{2}%
\beta_{\pm}/c^{3}$, $d\mathbf{k}^{\prime}=d\epsilon_{\gamma}^{\prime}%
\epsilon_{\gamma}^{\prime2}do_{\gamma}^{\prime}/c^{3}$ and the transition
function $W_{\mathbf{k}^{\prime},\mathbf{p}^{\prime};\mathbf{k},\mathbf{p}}$
is related to the transition probability differential $dw_{\mathbf{k}^{\prime
},\mathbf{p}^{\prime};\mathbf{k},\mathbf{p}}$ per unit time as%
\begin{equation}
W_{\mathbf{k}^{\prime},\mathbf{p}^{\prime};\mathbf{k},\mathbf{p}}%
d\mathbf{k}^{\prime}d\mathbf{p}^{\prime}\equiv Vdw_{\mathbf{k}^{\prime
},\mathbf{p}^{\prime};\mathbf{k},\mathbf{p}}. \label{TF}%
\end{equation}
The differential probability $dw_{\mathbf{k}^{\prime},\mathbf{p}^{\prime
};\mathbf{k},\mathbf{p}}=w_{\mathbf{k}^{\prime},\mathbf{p}^{\prime}%
;\mathbf{k},\mathbf{p}}d\mathbf{k}^{\prime}d\mathbf{p}^{\prime}$ is given by
(\ref{df-abs}) in Appendix C.

Given the momentum conservation one can perform one integration over
$d\mathbf{p}^{\prime}$ in (\ref{comptonemission}) as%
\begin{equation}
\int d\mathbf{p}^{\prime}\delta(\mathbf{k}+\mathbf{p}-\mathbf{k}^{\prime
}-\mathbf{p}^{\prime})\rightarrow1,
\end{equation}
but it is necessary to take into account the momentum conservation in the next
integration over $d\mathbf{k}^{\prime}$, so we have
\begin{gather}
\int d\epsilon_{\gamma}^{\prime}\delta(\epsilon_{\gamma}+\epsilon_{\pm
}-\epsilon_{\gamma}^{\prime}-\epsilon_{\pm}^{\prime})=\\
=\int d(\epsilon_{\gamma}^{\prime}+\epsilon_{\pm}^{\prime})\frac{1}%
{|\partial(\epsilon_{\gamma}^{\prime}+\epsilon_{\pm}^{\prime})/\partial
\epsilon_{\gamma}^{\prime}|}\delta(\epsilon_{\gamma}+\epsilon_{\pm}%
-\epsilon_{\gamma}^{\prime}-\epsilon_{\pm}^{\prime})\rightarrow\nonumber\\
\rightarrow\frac{1}{|\partial(\epsilon_{\gamma}^{\prime}+\epsilon_{\pm
}^{\prime})/\partial\epsilon_{\gamma}^{\prime}|}\equiv J_{\mathrm{cs}%
},\nonumber
\end{gather}
where the Jacobian of the transformation is%
\begin{equation}
J_{\mathrm{cs}}=\frac{1}{1-\beta_{\pm}^{\prime}\mathbf{b}_{\gamma}^{\prime
}\mathbf{\cdot b}_{\pm}^{\prime}},
\end{equation}
and $\mathbf{b}_{i}=\mathbf{p}_{i}/p$, $\mathbf{b}_{i}^{\prime}=\mathbf{p}%
_{i}^{\prime}/p^{\prime}$, $\mathbf{b}_{\pm}^{\prime}=(\beta_{\pm}%
\epsilon_{\pm}\mathbf{b}_{\pm}+\epsilon_{\gamma}\mathbf{b}_{\gamma}%
-\epsilon_{\gamma}^{\prime}\mathbf{b}_{\gamma}^{\prime})/(\beta_{\pm}^{\prime
}\epsilon_{\pm}^{\prime})$.

Finally, for the absorption coefficient we have
\begin{equation}
\chi^{\mathrm{cs}}f_{\gamma}=-\int do_{\gamma}^{\prime}d\mathbf{p}%
\frac{\epsilon_{\gamma}^{\prime}|M_{fi}|^{2}\hbar^{2}c^{2}}{16\epsilon_{\pm
}\epsilon_{\gamma}\epsilon_{\pm}^{\prime}}J_{\mathrm{cs}}f_{\gamma}%
(\mathbf{k},t)f_{\pm}(\mathbf{p},t), \label{comptonabscoef}%
\end{equation}
where the matrix element here is dimensionless. This integral is evaluated
numerically as described in Appendix.

For all binary interactions we use exact QED matrix elements which can be
found in the standard textbooks, e.g. in \cite{1982els..book.....B}%
,\cite{2003spr..book.....G},\cite{1981nau..book.....A}, and are given in
Appendix \ref{2body}.

In order to account for the charge screening we introduced the minimal
scattering angles following \cite{1988A&A...191..181H}, see Section
\ref{cutoff}\ in Appendix. This allows to apply the same scheme for the
computation of emission and absorption coefficients for Coulomb scattering,
while many treatments in the literature use the Fokker-Planck approximation,
e.g. \cite{1997ApJ...486..903P}.

For such a dense plasma collisional integrals in (\ref{BE})\ should include
not only binary interactions, having order $\alpha^{2}$ in Feynmann diagrams,
but also triple ones, having order $\alpha^{3}$ \cite{1982els..book.....B}. As
example for triple interactions consider relativistic bremsstrahlung
\begin{equation}
e_{1}+e_{2}\leftrightarrow e_{1}^{\prime}+e_{2}^{\prime}+\gamma^{\prime}.
\label{brems}%
\end{equation}
For the time derivative, for instance, of the distribution function $f_{2}$ in
the direct and in the inverse reactions (\ref{brems}) one has
\begin{gather}
\dot{f}_{2}=\int d\mathbf{p}_{1}d\mathbf{p}_{1}^{\prime}d\mathbf{p}%
_{2}^{\prime}d\mathbf{k}^{\prime}\left[  W_{\mathbf{p}_{1}^{\prime}%
,\mathbf{p}_{2}^{\prime},\mathbf{k}^{\prime};\mathbf{p}_{1},\mathbf{p}_{2}%
}f_{1}^{\prime}f_{2}^{\prime}f_{k}^{\prime}\right.  -\nonumber\\
-\left.  W_{\mathbf{p}_{1},\mathbf{p}_{2};\mathbf{p}_{1}^{\prime}%
,\mathbf{p}_{2}^{\prime},\mathbf{k}^{\prime}}f_{1}f_{2}\right]  =\int
d\mathbf{p}_{1}d\mathbf{p}_{1}^{\prime}d\mathbf{p}_{2}^{\prime}d\mathbf{k}%
^{\prime}\frac{c^{6}\hbar^{3}}{(2\pi)^{2}}\times\\
\times\frac{\delta^{(4)}(P_{f}-P_{i})|M_{fi}|^{2}}{2^{5}\epsilon_{1}%
\epsilon_{2}\epsilon_{1}^{\prime}\epsilon_{2}^{\prime}\epsilon_{\gamma
}^{\prime}}\left[  f_{1}^{\prime}f_{2}^{\prime}f_{k}^{\prime}-\frac{1}%
{(2\pi\hbar)^{3}}f_{1}f_{2}\right]  ,\nonumber
\end{gather}
where%
\begin{align*}
d\mathbf{p}_{1}d\mathbf{p}_{2}W_{\mathbf{p}_{1}^{\prime},\mathbf{p}%
_{2}^{\prime},\mathbf{k}^{\prime};\mathbf{p}_{1},\mathbf{p}_{2}}  &  \equiv
V^{2}dw_{1},\\
d\mathbf{p}_{1}^{\prime}d\mathbf{p}_{2}^{\prime}d\mathbf{k}^{\prime
}W_{\mathbf{p}_{1},\mathbf{p}_{2};\mathbf{p}_{1}^{\prime},\mathbf{p}%
_{2}^{\prime},\mathbf{k}^{\prime}}  &  \equiv Vdw_{2},
\end{align*}
and $dw_{1}$ and $dw_{2}$\ are given by (\ref{dp}) for the inverse and direct
process (\ref{brems}) respectively. The matrix element here has dimensions of
the length squared, see Section \ref{ME} in Appendix.

In the case of the distribution functions (\ref{dk}), see below, we have
multipliers proportional to
\begin{equation}
F_{i}=\exp\frac{\nu_{i}}{\theta_{i}}, \label{fugacity}%
\end{equation}
called fugacities, in front of the integrals. The calculation of emission and
absorption coefficients is then reduced to the well known thermal equilibrium
case \cite{1984MNRAS.209..175S}. In fact, since reaction rates of triple
interactions are $\alpha$ times smaller than binary reaction rates, we expect
that binary reactions come to detailed balance first. Only when binary
reactions are all balanced, triple interactions become important. In addition,
when binary reactions come into balance, distribution functions already
acquire the form (\ref{dk}). Although there is no principle difficulty in
computations using exact matrix elements for triple reactions as well, our
simplified scheme allows for much faster numerical computation. The
corresponding reaction rates for triple interactions are given is Section
\ref{3body} in Appendix.

We consider all possible binary and triple interactions between electrons,
positrons and photons as summarized in table~\ref{tab1}.%

\begin{table}[tbp] \centering
\begin{tabular}
[c]{|c|c|}\hline\hline
Binary interactions & Radiative and\\
& pair producing variants\\\hline\hline
{M{\o }ller and Bhabha} & {Bremsstrahlung}\\
{$e_{1}^{\pm}{e_{2}^{\pm}\longrightarrow e_{1}^{\pm}}^{\prime}$}${e_{2}^{\pm}%
}^{\prime}$ & {$e_{1}^{\pm}e_{2}^{\pm}{\leftrightarrow}e_{1}^{\pm\prime}%
e_{2}^{\pm\prime}\gamma$}\\
{$e^{\pm}{e^{\mp}\longrightarrow e^{\pm\prime}}$}${e^{\mp\prime}}$ & {$e^{\pm
}e^{\mp}{\leftrightarrow}e^{\pm\prime}e{^{\mp\prime}}\gamma$}\\\hline
Single {Compton} & {Double Compton}\\
{\ $e^{\pm}\gamma{\longrightarrow}e^{\pm}\gamma^{\prime}$} & {$e^{\pm}%
\gamma{\leftrightarrow}e^{\pm\prime}\gamma^{\prime}\gamma^{\prime\prime}$%
}\\\hline
{Pair production} & Radiative pair production\\
and annihilation & and 3-photon annihilation\\
{$\gamma\gamma^{\prime}{\leftrightarrow}e^{\pm}e^{\mp}$} & $\gamma
\gamma^{\prime}${${\leftrightarrow}e^{\pm}e^{\mp}$}$\gamma^{\prime\prime}$\\
& {$e^{\pm}e^{\mp}{\leftrightarrow}\gamma\gamma^{\prime}$}$\gamma
^{\prime\prime}$\\\hline
& $e^{\pm}\gamma${${\leftrightarrow}e^{\pm\prime}{e^{\mp}}e^{\pm\prime\prime}
$}\\\hline\hline
\end{tabular}
\caption{Microphysical processes in the pair plasma.}\label{tab1}%
\end{table}
Each of the above mentioned reactions is characterized by the corresponding
timescale and optical depth. For Compton scattering of an electron, for
instance, we have%
\begin{equation}
t_{\mathrm{cs}}=\frac{1}{\sigma_{T}n_{\pm}c},\qquad\tau_{\mathrm{cs}}%
=\sigma_{T}n_{\pm}R, \label{ttau}%
\end{equation}
where $\sigma_{T}=\frac{8\pi}{3}\alpha^{2}(\frac{\hbar}{mc})^{2}$ is the
Thomson cross-section. There are two timescales in our problem that
characterize the condition of detailed balance between direct and inverse
reactions, $t_{\mathrm{cs}}$ for binary and $\alpha^{-1}t_{\mathrm{cs}}$ for
triple interactions respectively.

We choose arbitrary initial distribution functions and find a common
development. At a certain time $t_{\mathrm{k}}$ the distribution functions
always have evolved in a functional form on the entire energy range, and
depend only on two parameters. We find in fact for the distribution functions
the expressions
\begin{equation}
f_{i}(\varepsilon)=\frac{2}{(2\pi\hbar)^{3}}\exp\left(  -\frac{\varepsilon
-\nu_{i}}{\theta_{i}}\right)  , \label{dk}%
\end{equation}
with chemical potential $\nu_{i}\equiv\frac{\varphi_{i}}{mc^{2}}$ and
temperature $\theta_{i}\equiv\frac{k_{B}T_{i}}{m_{e}c^{2}}$, where
$\varepsilon\equiv\frac{\epsilon}{m_{e}c^{2}}$ is the energy of the particle.
Such a configuration corresponds to a kinetic equilibrium
\cite{1990eaun.book.....K},\cite{1997ApJ...486..903P}%
,\cite{1973rela.conf....1E} in which particles acquire a common temperature
and nonzero chemical potentials. At the same time we found that triple
interactions become essential for $t>t_{\mathrm{k}}$, after the establishment
of kinetic equilibrium. In strict mathematical sense the sufficient condition
for reaching thermal equilibrium is when all direct reactions are exactly
balanced with their inverse. Therefore, in principle, not only triple, but
also four-particle, five-particle and so on reaction have to be accounted for
in equation (\ref{BE}). The timescale for reaching thermal equilibrium will be
then determined by the slowest reaction which is not balanced with its
inverse. We stress, however, that the necessary condition is the detailed
balance at least in triple interactions, since binary reactions do not change
chemical potentials.

Notice that a method similar to ours was applied in \cite{1997ApJ...486..903P}
in order to compute spectra of particles in kinetic equilibrium. However,
although the approach was similar, the computation was never carried out in
order to actually observe the reaching of thermal equilibrium.

Finally, it is worth mentioning the physical meaning of the chemical potential
$\nu_{\mathrm{k}}$ in kinetic equilibrium entering the formula (\ref{dk}). In
the case of pure pair plasma a non-zero chemical potential represents
deviation from the thermal equilibrium through the relation%
\begin{equation}
\nu_{\mathrm{k}}=\theta\ln(n_{\mathrm{k}}/n_{\mathrm{th}}),
\end{equation}
where $n_{\mathrm{th}}$ are concentrations of particles in thermal equilibrium.

\section{Proton loading}

So far we dealt with leptons, having the same mass but opposite charges. In
that case the condition of electric neutrality is identically fulfilled. We
described electrons and positrons with the same distribution function.
Situation becomes more complicated when admixture of protons is allowed. Since
charge neutrality%
\begin{equation}
n_{-}=n_{+}+n_{p} \label{chargecons}%
\end{equation}
is required, the number of electrons is not equal to the number of protons. In
such a case a new dimensionless parameter, the baryonic loading $\mathbf{B}$,
can be introduced as%
\begin{equation}
\mathbf{B}=\frac{NMc^{2}}{\mathcal{E}}=\frac{n_{p}Mc^{2}}{\rho_{r}}, \label{B}%
\end{equation}
where $N$ and $n_{p}$ are the number and the concentration of protons,
$\mathcal{E}$ and $\rho_{r}=\rho_{\gamma}+\rho_{+}+\rho_{-}$ are radiative
energy and energy density respectively. Since in relativistic plasma electrons
and positrons move with almost the speed of light, both photons and pairs in
thermal equilibrium behave as relativistic fluid with equation of state
$p_{r}\simeq\rho_{r}/3$. At the same time, protons are relatively particles in
the energy range (\ref{avenergy}), with negligible pressure and dust-like
equation of state $p\simeq0$. In this way by introducing parameter
$\mathbf{B}$ we distinguish a radiation-dominated ($\mathbf{B}<1$) from a
matter-dominated ($\mathbf{B}>1$) plasma. For electrically neutral plasmas
there exists an upper limit on the parameter $\mathbf{B}$\ defined by
(\ref{B}), which is $\mathbf{B\leq}M/m$.

In the range of energies (\ref{avenergy}) the radiative energy density can be
approximated as $\rho_{r}\sim n_{-}mc^{2}$, and then we have for
concentrations $n_{p}\sim n_{-}\mathbf{B}\frac{m}{M}$. If protons and
electrons are at the same temperature then from the equality of the kinetic
energy of a proton $\epsilon_{k,p}=\frac{Mv_{p}^{2}}{2}$ and the one of an
electron $\epsilon_{k,-}\sim mc^{2}$ we have $\frac{v_{p}}{c}\sim\sqrt
{\frac{m}{M}}$, therefore protons are indeed nonrelativistic.%

\begin{table}[tbp] \centering
\begin{tabular}
[c]{|c|c|}\hline\hline
Binary interactions & Radiative and\\
& pair producing variants\\\hline\hline
Coulomb scattering & {Bremsstrahlung}\\
$p${$_{1}{p_{2}\longrightarrow p}_{1}^{\prime}{p}_{2}^{\prime}$} & $p$%
{$_{1}{p_{2}\leftrightarrow p}_{1}^{\prime}{p}_{2}^{\prime}$}$\gamma$\\
$p${${e^{\pm}\longrightarrow p}^{\prime}e^{\pm\prime}$} & $p${${e^{\pm
}\leftrightarrow p}^{\prime}e^{\pm\prime}$}$\gamma$\\\hline
& $p${${e_{1}^{\pm}\leftrightarrow p}^{\prime}e_{1}^{\pm\prime}$}$e^{\pm
}e^{\mp}$\\\hline
Single {Compton} & {Double Compton}\\
$p${$\gamma{\longrightarrow}p^{\prime}\gamma^{\prime}$} & $p${$\gamma
{\leftrightarrow}p^{\prime}\gamma^{\prime}\gamma^{\prime\prime}$}\\\hline
& $p${$\gamma{\leftrightarrow}p^{\prime}$}$e^{\pm}e^{\mp}$\\\hline\hline
\end{tabular}
\caption{Microphysical processes in the pair plasma involving protons.}\label{tab1a}%
\end{table}%

In presence of protons additional binary reactions consist of Coulomb
collisions between electrons (positrons) and protons, scattering of protons on
protons and Compton scattering of protons. Additional triple reactions are
radiative variants of these reactions, see Table \ref{tab1a} and Appendix
\ref{2body_p}.

Protons can be thermalized in two ways:\ either in a two-step process first
between themselves and then by electron/positron-proton collisions, or just by
the latter mechanism. The rate of proton-proton collisions is a factor
$\sqrt{\frac{m}{M}}\frac{n_{p}}{n_{-}}\sim\mathbf{B}\left(  \frac{m}%
{M}\right)  ^{3/2}$ smaller than the rate of electron-electron collisions, see
(\ref{ppR}). The rate of proton-electron/positron collisions is a factor
$\frac{\epsilon}{Mc^{2}}\sim\frac{m}{M}$ smaller than the one of
electron-electron collisions, see (\ref{epR}). Therefore, for $\mathbf{B}%
>\sqrt{\frac{m}{M}}$ proton-proton collisions are faster, while for
$\mathbf{B}<\sqrt{\frac{m}{M}}$ proton-electron/positron ones predominate.

\section{The discretization procedure and the computational scheme}

\label{discretization}

In order to solve equations (\ref{BE}) we use a finite difference method by
introducing a computational grid in the phase space to represent the
distribution functions and to compute collisional integrals following
\cite{2004ApJ...609..363A}. Our goal is to construct the scheme implementing
energy, baryon number and electric charge conservation laws, see Appendix
\ref{conslaws}. For this reason we prefer to use in the code, instead of
distribution functions $f_{i}$, the spectral energy densities
\begin{equation}
E_{i}(\epsilon_{i})=\frac{4\pi\epsilon_{i}^{3}\beta_{i}f_{i}}{c^{3}},
\label{Efuncs}%
\end{equation}
where $\beta_{i}=\sqrt{1-(m_{i}c^{2}/\epsilon_{i})^{2}}$, in the phase space
$\epsilon_{i}$. Then
\begin{equation}
\epsilon_{i}f_{i}(\mathbf{p},t)d\mathbf{r}d\mathbf{p}=\frac{4\pi\epsilon
^{3}\beta_{i}f_{i}}{c^{3}}d\mathbf{r}d\epsilon_{i}=E_{i}d\mathbf{r}%
d\epsilon_{i}%
\end{equation}
is the energy in the volume of the phase space $d\mathbf{r}d\mathbf{p}$. The
number density of particles of type "$i$" is given by%
\begin{equation}
n_{i}=\int f_{i}d\mathbf{p}=\int\frac{E_{i}}{\epsilon_{i}}d\epsilon_{i},\qquad
dn_{i}=f_{i}d\mathbf{p},
\end{equation}
while the corresponding energy density is%
\[
\rho_{i}=\int\epsilon_{i}f_{i}d\mathbf{p}=\int E_{i}d\epsilon_{i}.
\]

We can rewrite Boltzmann equations (\ref{BE}) in the form
\begin{equation}
\frac{1}{c}\frac{\partial E_{i}}{\partial t}=\sum_{q}(\tilde{\eta}_{i}%
^{q}-\chi_{i}^{q}E_{i}), \label{EBoltzmannEq}%
\end{equation}
where $\tilde{\eta}_{i}^{q}=(4\pi\epsilon_{i}^{3}\beta_{i}/c^{3})\eta_{i}^{q}
$.

We introduced the computational grid for phase space $\{\epsilon_{i},\mu
,\phi\}$, where $\mu=\cos\vartheta$, $\vartheta$ and $\phi$ are angles between
radius vector $\mathbf{r}$ and the particle momentum $\mathbf{p}$. The zone
boundaries are $\epsilon_{i,\omega\mp1/2}$, $\mu_{k\mp1/2}$, $\phi_{l\mp1/2}$
for $1\leq\omega\leq\omega_{\mathrm{max}}$, $1\leq k\leq k_{\mathrm{max}}$,
$1\leq l\leq l_{\mathrm{max}}$. The length of the $i $-th interval is
$\Delta\epsilon_{i,\omega}\equiv\epsilon_{i,\omega+1/2}-\epsilon
_{i,\omega-1/2}$.\ On the finite grid the functions (\ref{Efuncs}) become
\begin{equation}
E_{i,\omega}\equiv\frac{1}{\Delta\epsilon_{i,\omega}}\int_{\Delta
\epsilon_{i,\omega}}d\epsilon E_{i}(\epsilon).
\end{equation}

Now we can replace the collisional integrals in (\ref{EBoltzmannEq}) by the
corresponding sums.

After this procedure we get the set of ordinary differential equations
(ODE's), instead of the system of partial differential equations for the
quantities $E_{i,\omega}$ to be solved. There are several characteristic times
for different processes in the problem, and therefore our system of
differential equations is stiff. Under these conditions eigenvalues of Jacobi
matrix differs significantly, and the real parts of eigenvalues are negative.
We use Gear's method \cite{1976oup..book.....H} to integrate ODE's
numerically. This high-order implicit method was developed for the solution of
stiff ODE's.

In our method exact energy conservation law is satisfied. For binary
interactions the particles number conservation law is satisfied as we adopt
interpolation of grid functions $E_{i,\omega}$ inside the energy intervals.

\section{Numerical results}

In what follows we consider in details three specific cases. In the first two
cases our grid consists of 60 energy intervals and $16\times32$ intervals for
two angles $\vartheta$ and $\phi$\ characterizing the direction of the
particle momentum. In the third case we have 40 energy intervals.

\subsection{Case I}

We take the following initial conditions: flat initial spectral densities
$E_{i}(\epsilon_{i})=\mathrm{const}$, total energy density $\rho
=10^{24}\mathrm{erg}/\mathrm{cm}^{3}$. Plasma is dominated by photons with
small amount of electron-positron pairs, the ratio between energy densities in
photons and in electron-positron pairs $\zeta=\rho_{\pm}/\rho_{\gamma}%
=10^{-5}$. Baryonic loading parameter $\mathbf{B}=10^{-3}$, corresponding to
$\rho_{p}=2.7\times10^{18}\mathrm{erg}/\mathrm{cm}^{3}$.
\begin{figure}
[h]
\begin{center}
\includegraphics[
height=2.6161in,
width=3.6201in
]%
{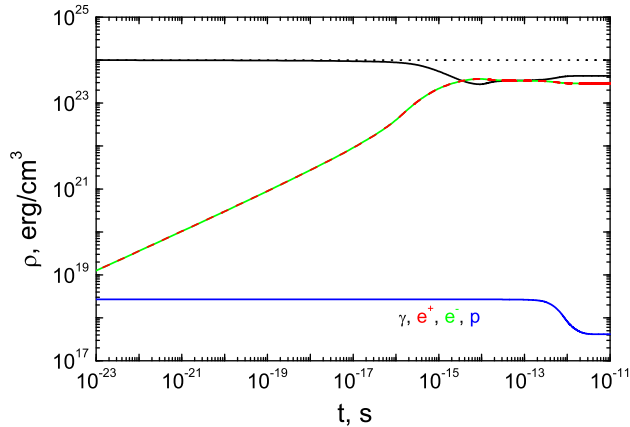}%
\caption{Depencence on time of energy densities of electrons (green),
positrons (red), photons (black) and protons (blue) for initial conditions I.
Total energy density is shown by dotted black line. Interaction between pairs
and photons operates on very short timescales up to $10^{-23}$ sec.
Quasi-equilibrium state is established at $t_{\mathrm{k}}\simeq10^{-14}$ sec
which corresponds to kinetic equilibrium for pairs and photons. Protons start
to interact with then as late as at $t_{\mathrm{th}}\simeq10^{-13}$ sec.}%
\label{rho1}%
\end{center}
\end{figure}
\begin{figure}
[h]
\begin{center}
\includegraphics[
height=2.5746in,
width=3.6322in
]%
{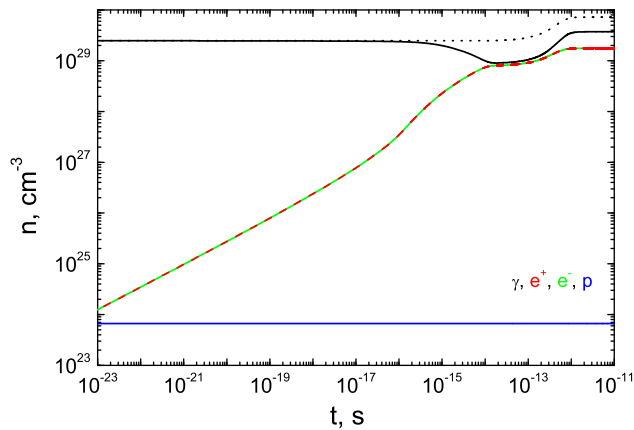}%
\caption{Depencence on time of concentrations of electrons (green), positrons
(red), photons (black) and protons (blue) for initial conditions I. Total
number density is shown by dotted black line. In this case kinetic equilibrium
between electrons, positrons and photons is reached at $t_{\mathrm{k}}%
\simeq10^{-14}$ sec. Protons join thermal equilibrium with other particles at
$t_{\mathrm{th}}\simeq4\times10^{-12}$ sec.}%
\label{n1}%
\end{center}
\end{figure}
\begin{figure}
[h]
\begin{center}
\includegraphics[
height=2.6256in,
width=3.4843in
]%
{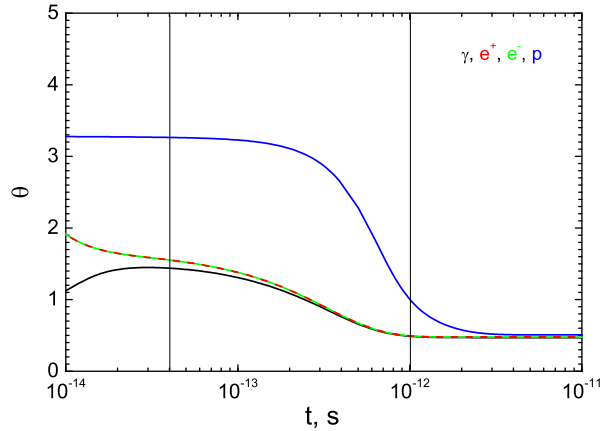}%
\caption{Depencence on time of dimensionless temperature of electrons (green),
positrons (red), photons (black) and protons (blue) for initial conditions I.
The temperature for pairs and photons acquires physical meaning only in
kinetic equilibrium at $t_{\mathrm{k}}\simeq10^{-14}$ sec. Protons are cooled
by the pair-photon plasma and acquire common temperature with it as late as at
$t_{\mathrm{th}}\simeq4\times10^{-12}$ sec.}%
\label{theta1}%
\end{center}
\end{figure}
\begin{figure}
[h]
\begin{center}
\includegraphics[
height=2.5538in,
width=3.5786in
]%
{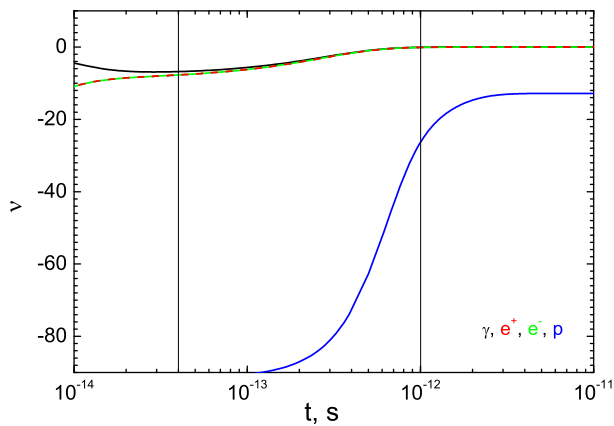}%
\caption{Depencence on time of dimensionless chemical potential of electrons
(green), positrons (red), photons (black) and protons (blue) for initial
conditions I. The chemical potential for pairs and photons acquires physical
meaning only in kinetic equilibrium at $t_{\mathrm{k}}\simeq10^{-14}$ sec,
while for protons this happens at $t_{\mathrm{th}}\simeq4\times10^{-12}$ sec.
At this time chemical potential of photons has evolved to zero and thermal
equilibrium has been already reached.}%
\label{phi1}%
\end{center}
\end{figure}
\begin{figure}
[h]
\begin{center}
\includegraphics[
height=4.3422in,
width=3.3745in
]%
{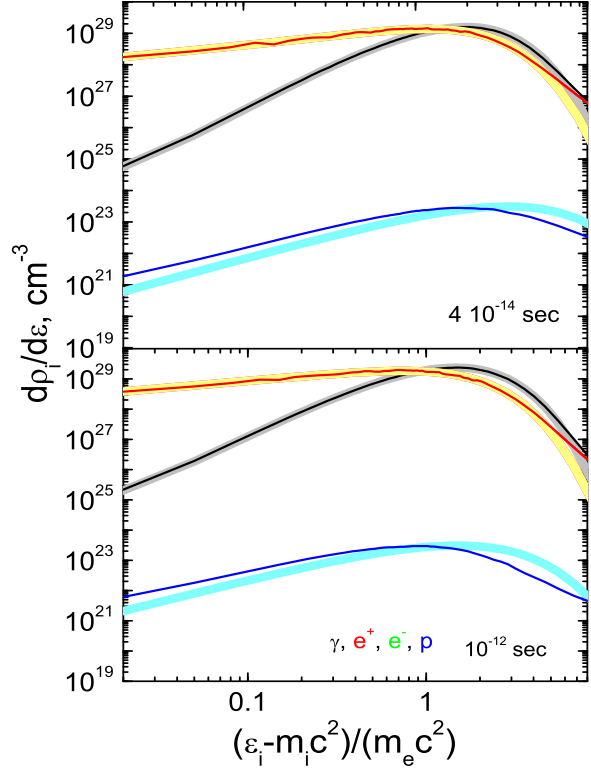}%
\caption{Spectral density as function of particle energy for electrons
(green), positrons (red), photons (black) and protons (blue) for initial
conditions I at intermediate time moments $t_{1}=4\times10^{-14}$ sec (upper
figure) and $t_{2}=10^{-12}$ sec (lower figure). Fits of the spectra with
chemical potentials and temperatures corresponding to thermal equilibrium
state are also shown by yellow (electrons and positrons), grey (photons) and
light blue (protons) thick lines. The upper figure shows the spectra when
kinetic equilibrium is established for the first time between electrons,
positrons and photons while the lower figure shows the spectra at thermal
equilibrium between these particles. On both figures protons are not yet in
equilibrium neither with themselves nor with other particles.}%
\label{spectra1a}%
\end{center}
\end{figure}
\begin{figure}
[h]
\begin{center}
\includegraphics[
height=2.5841in,
width=3.5172in
]%
{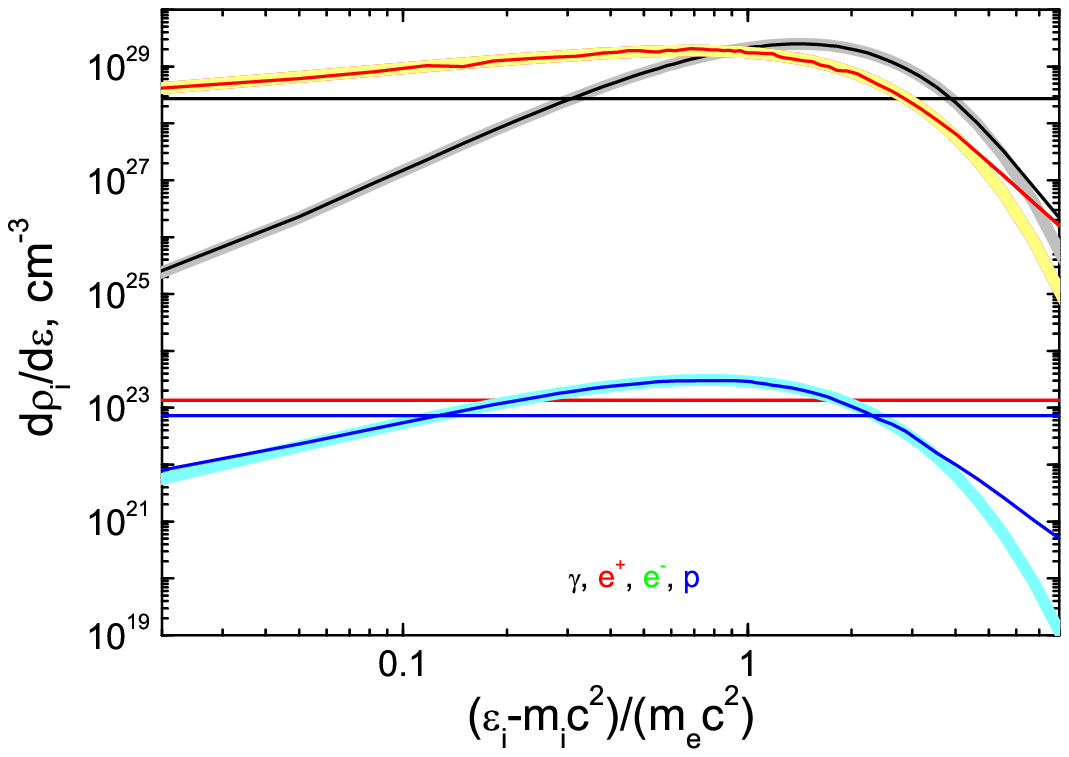}%
\caption{Spectral density as function of particle energy are shown as before
at initial and final moments of the computations. The final photon spectrum is
black body one.}%
\label{spectra1}%
\end{center}
\end{figure}
The energy density in each component of plasma changes, as can be seen from
fig. \ref{rho1}, keeping constant the total energy density shown by dotted
line in fig. \ref{rho1}, as the energy conservation requires. As early as at
$10^{-23}$ sec the energy starts to redistribute between electrons and
positrons from the one hand and photons from the other hand essentially by the
pair-creation process. This leads to equipartition of energies between these
particles at $3\times10^{-15}$\ sec. Concentrations of pairs and photons
equalize at $10^{-14}$\ sec, as can be seen from fig. \ref{n1}. From this
moment temperatures and chemical potentials of electrons, positrons and
photons tend to be equal, see fig. \ref{theta1} and fig. \ref{phi1}
respectively, and it corresponds to the approach to kinetic equilibrium.

This is quasi-equilibrium state since total number of particles is still
approximately conserved, as can be seen from fig. \ref{n1}, and triple
interactions are not yet efficient. At the moment $t_{1}=4\times10^{-14}%
$\ sec, shown by the vertical line on the left in fig. \ref{theta1} and fig.
\ref{phi1}, the temperature of photons and pairs is $\theta_{\mathrm{k}}%
\simeq1.5$, while the chemical potentials of these particles are
$\nu_{\mathrm{k}}\simeq-7$. Concentration of protons is so small that their
energy density is not affected by the presence of other components; also
proton-proton collisions are inefficient. In other words, protons do not
interact yet and their spectra are not yet of equilibrium form, see fig.
\ref{spectra1a}. The temperature of protons start to change only at $10^{-13}%
$\ sec, when proton-electron Coulomb scattering becomes efficient.

As can be seen from fig. \ref{phi1}, the chemical potentials of electrons,
positrons and photons evolved by that time due to triple interactions. Since
chemical potentials of electrons, positrons and photons were negative, the
particles were in deficit with respect to the thermal state. This caused the
total number of these particles to increase and consequently the temperature
to decrease. The chemical potential of photons reaches zero at $t_{2}%
=10^{-12}$\ sec, shown by the vertical line on the right in fig. \ref{theta1}
and fig. \ref{phi1}, which means that electrons, positrons and photons are now
in thermal equilibrium. However, protons are not yet in equilibrium with other
particle since their spectra are not thermal, as shown in the lower part of
fig. \ref{spectra1a}.

Finally, the proton component thermalize with other particles at
$4\times10^{-12}$\ sec, and from that moment plasma is characterized by unique
temperature, $\theta_{\mathrm{th}}\simeq0.48$ as fig. \ref{theta1} clearly
shows. Protons have final chemical potential $\nu_{p}\simeq-12.8$.

This state is characterized by thermal distribution of all particles as can be
seen from fig. \ref{spectra1}. There initial flat as well as final spectral
densities are shown together with fits of particles spectra with the values of
the common temperature and the corresponding chemical potentials in thermal equilibrium.

\subsection{Case II}

We take the following initial conditions: power law spectral densities
$E_{i}(\epsilon_{i})$ for protons, electrons and positrons with initial energy
densities $\rho_{p}=2.8\times10^{22}$ erg/cm$^{3}$, $\rho_{-}=1.5\times
10^{24}$ erg/cm$^{3}$, $\rho_{+}=1.5\times10^{21}$ erg/cm$^{3}$, respectively.
We chosen flat spectral density for photons with $\rho_{\gamma}=2.8\times
10^{24}$ erg/cm$^{3}$. Initial baryonic loading parameter is set to
$\mathbf{B}=608$, corresponding to a matter-dominated plasma, unlike the
previous case.
\begin{figure}
[h]
\begin{center}
\includegraphics[
height=2.5105in,
width=3.5674in
]%
{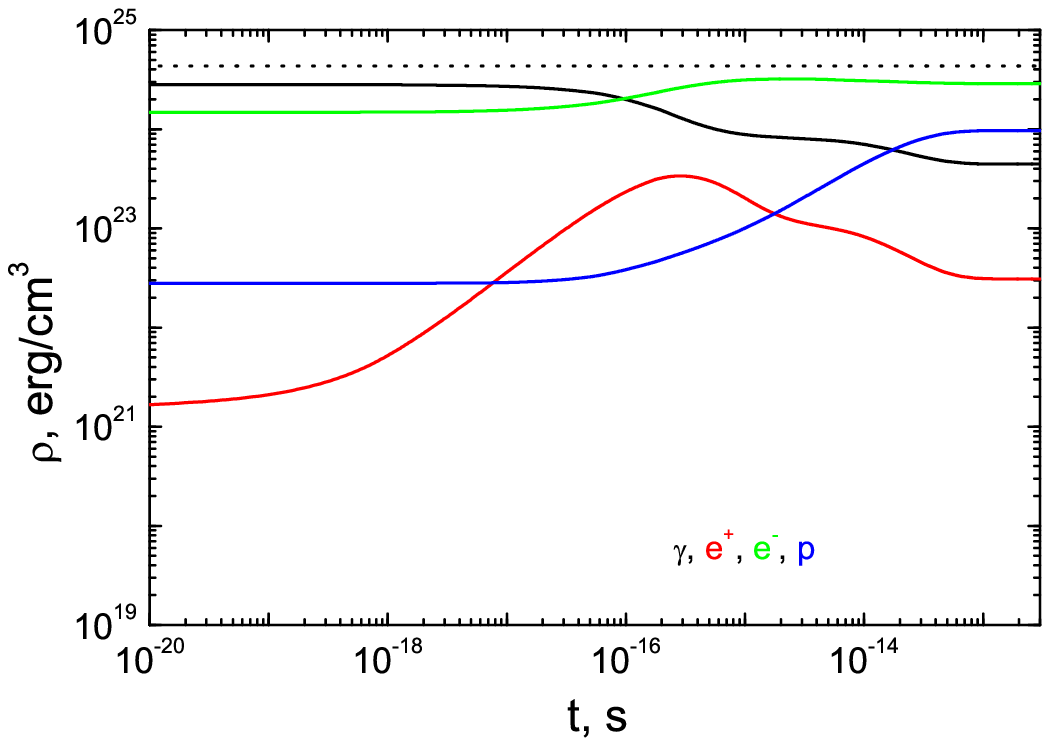}%
\caption{Depencence on time of energy densities for initial conditions II.
Colors are as in the case I. Protons start to interact with other particles as
late as at $t\simeq10^{-16}$ sec.}%
\label{rho2}%
\end{center}
\end{figure}
\begin{figure}
[h]
\begin{center}
\includegraphics[
height=2.6481in,
width=3.6322in
]%
{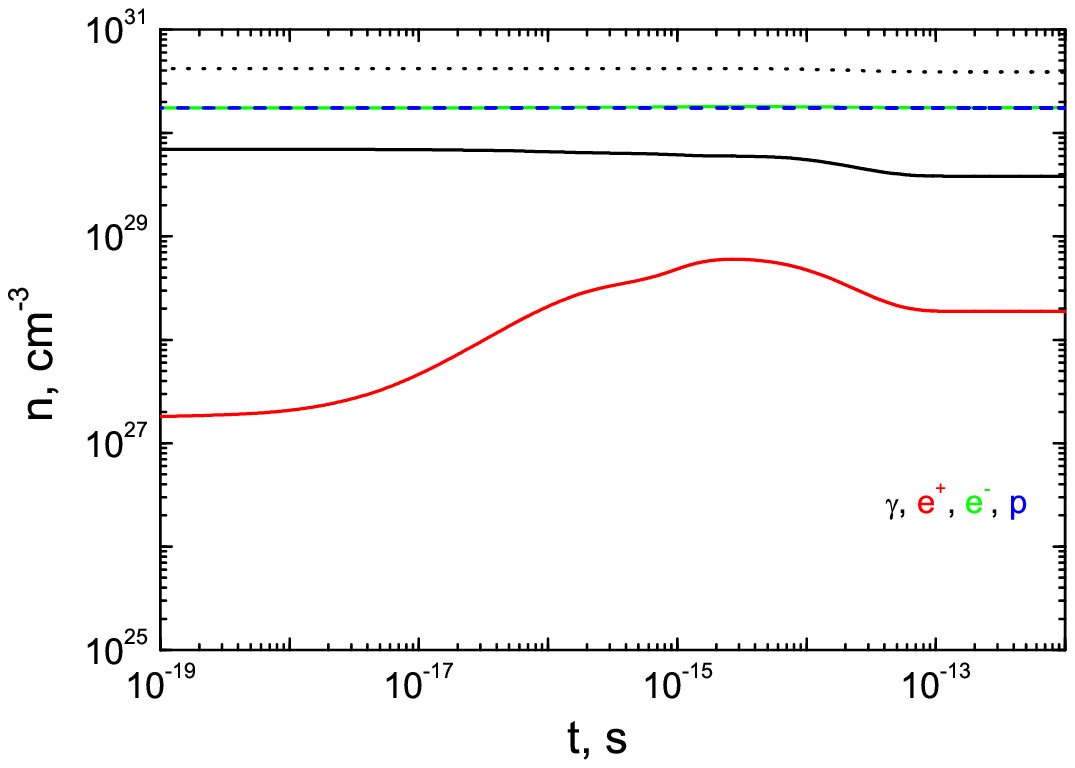}%
\caption{Depencence on time of concentrations for initial conditions II.
Colors are as in the case I.}%
\label{n2}%
\end{center}
\end{figure}
\begin{figure}
[h]
\begin{center}
\includegraphics[
height=2.6256in,
width=3.4843in
]%
{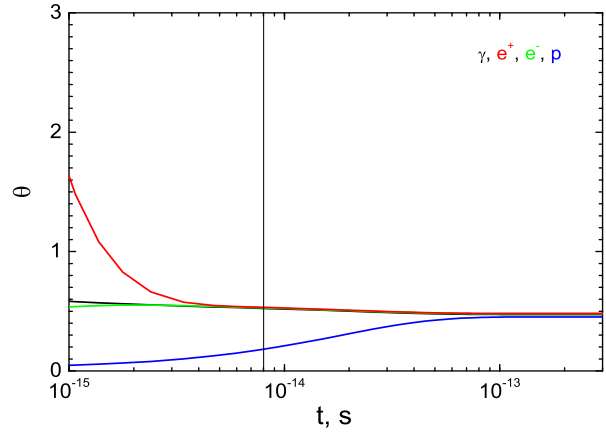}%
\caption{Depencence on time of dimensionless temperature for initial
conditions II. Colors are as in the case I. The pair-photon plasma is heating
protons. Protons join thermal equilibrium at $t_{\mathrm{th}}\simeq10^{-13}$
sec.}%
\label{theta2}%
\end{center}
\end{figure}
\begin{figure}
[h]
\begin{center}
\includegraphics[
height=2.5538in,
width=3.5786in
]%
{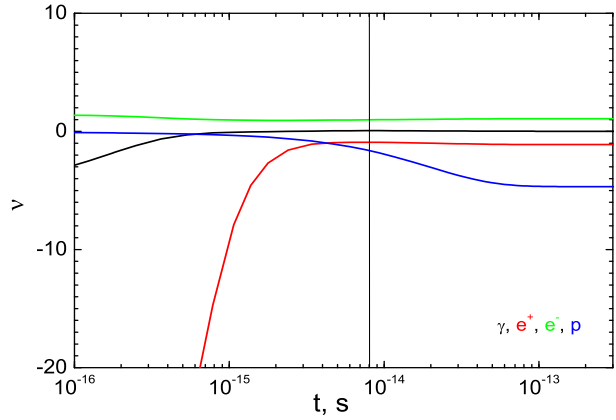}%
\caption{Depencence on time of dimensionless chemical potential for initial
conditions II. Colors are as in the case I. The chemical potential of photons
is almost zero in kinetic equilibrium. The chemical potentials of electrons
and positrons are almost equal and opposite in kinetic equilibrium, to
maintain electric neutrality.}%
\label{phi2}%
\end{center}
\end{figure}
\begin{figure}
[h]
\begin{center}
\includegraphics[
height=2.6896in,
width=3.5466in
]%
{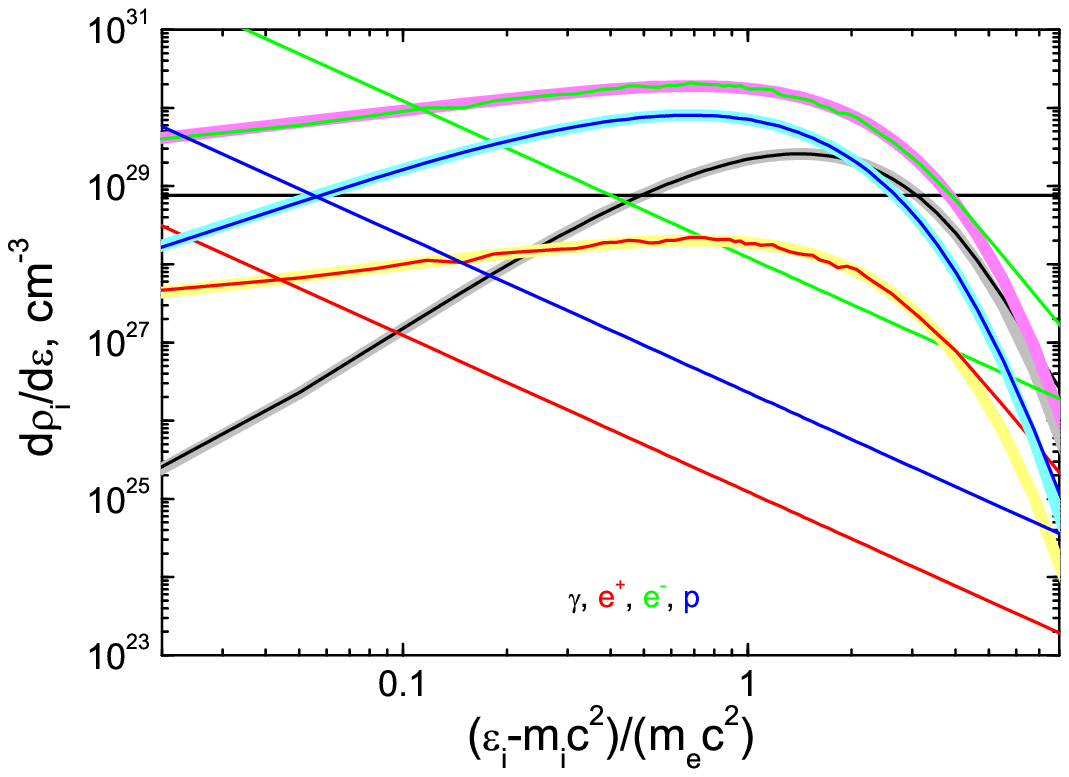}%
\caption{Initial and final spectral density as function of particle energy for
initial conditions II. Fits of the final spectra with chemical potentials and
temperatures are also shown.}%
\label{spectra2a}%
\end{center}
\end{figure}
As in the case I, the most rapid reaction is electron-positron pair creation
which starts to change the energy density of positrons at $10^{-20}$ sec, see
fig. \ref{rho2}. Initially most energy is in photons, followed by electrons
and protons. In the course of the evolution the energy gets redistributed in
such a way that in the final state most energy is transferred first to the
electrons, then follow the protons, the photons and finally the positrons. In
fig. \ref{n2} one can see that number densities of electrons and protons are
almost equal with chosen heavy proton loading. Concentrations of particles
almost do not change during evolution towards thermal equilibrium.

Temperatures and chemical potentials of particles are shown in fig.
\ref{theta2} and \ref{phi2} respectively. Kinetic equilibrium is established
at around $8\times10^{-15}$ sec, marked by the vertical line. The temperature
of pairs and photons at that moment is $\theta_{\mathrm{k}}\simeq0.53$, while
the chemical potentials of these particles are $\nu_{-}\simeq1$, $\nu
_{+}\simeq-0.9$, $\nu_{\gamma}\simeq0.1$. Notice that chemical potentials of
electrons and positrons are almost equal in magnitude and opposite in kinetic
equilibrium, see fig. \ref{phi2}. At this moment protons are not yet in
equilibrium with the rest of plasma but already established kinetic
equilibrium with themselves with the temperature $\theta_{p}\simeq0.18$ and
the chemical potential $\nu_{p}\simeq-2$. The common temperature is reached at
the moment $10^{-13}$ sec, which corresponds to thermal equilibrium. Final
values of temperature is $\theta_{\mathrm{th}}\simeq0.47 $, while chemical
potentials are $\nu_{\pm}\simeq\mp1$, $\nu_{p}\simeq-4.7$.

The share of the proton energy density in the total energy density increased
in course of time, see fig. \ref{rho2}, causing an increase in the baryonic
loading parameter which reached in thermal equilibrium the value
$\mathbf{B}=780$.

Since concentration of protons is chosen to be large, proton-proton collisions
become more important than proton-electron/positron collisions, in contrast to
the case I. In fact, protons reached equilibrium temperature already at
$10^{-16}$ sec, while they start to interact with electrons and positrons only
at $10^{-15}$ sec.

\subsection{Case III}

We take the following initial conditions: the initial ratio between
concentrations of electrons and protons is $\varsigma=n_{p}/n_{-}=10^{-3}$.
The total energy density is chosen in such a way that the final temperature in
thermal equilibrium be $\theta_{\mathrm{th}}=2$. We set up flat initial
spectrum for photons $E_{\gamma}(\epsilon_{i})=\mathrm{const}$, and power law
spectra for the pairs $E_{\pm}(\epsilon_{\pm})\propto\left[  \epsilon_{\pm
}-mc^{2}\right]  ^{-2}$ and protons $E_{p}(\epsilon_{p})\propto\left[
\epsilon_{p}-Mc^{2}\right]  ^{-4}$. Finally, the ratio of initial and final
concentrations of positrons is chosen to be $n_{+}=10^{-1}n_{+}^{\mathrm{th}}
$. Given these initial conditions the baryon loading parameter is
$\mathbf{B}=0.2$.
\begin{figure}
[h]
\begin{center}
\includegraphics[
height=2.5322in,
width=3.6002in
]%
{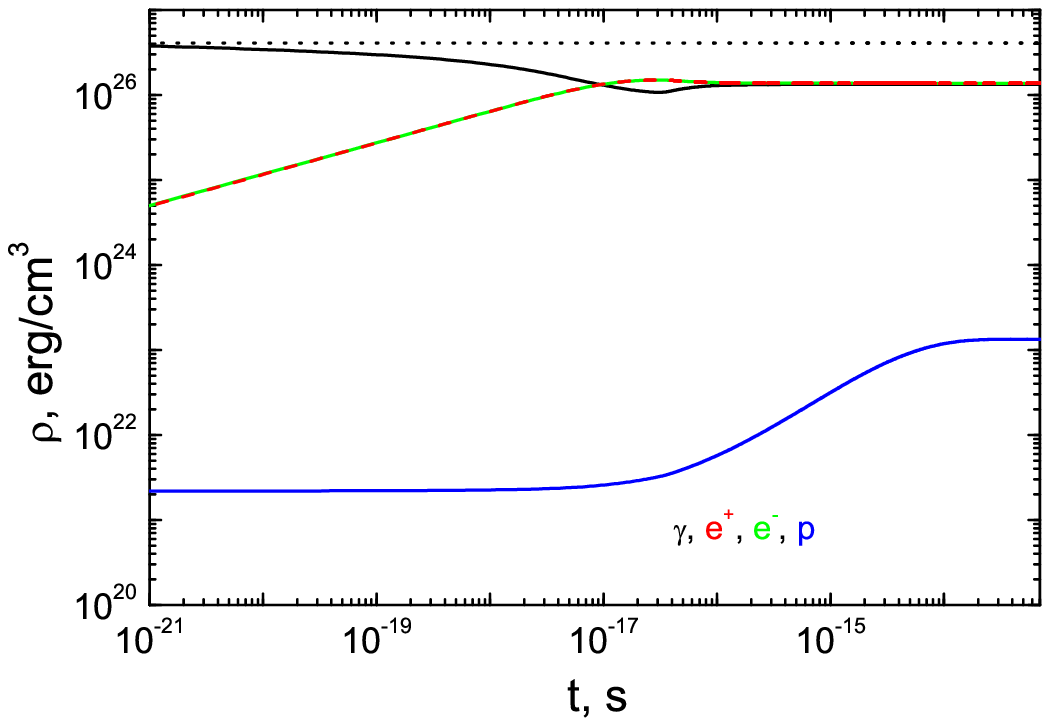}%
\caption{Depencence on time of energy densities for initial conditions III.
Colors are as in the case I. Protons start to interact with other particles at
about $10^{-17}$ sec.}%
\label{rho3}%
\end{center}
\end{figure}
\begin{figure}
[h]
\begin{center}
\includegraphics[
height=2.6896in,
width=3.557in
]%
{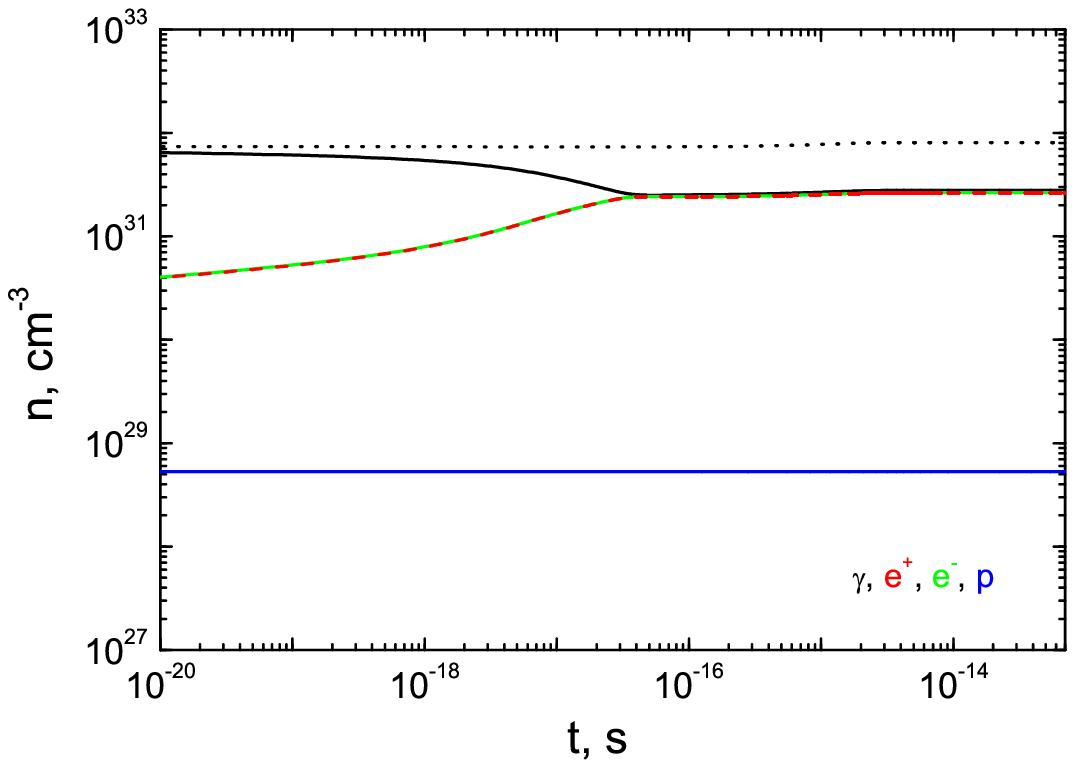}%
\caption{Depencence on time of concentrations for initial conditions III.
Colors are as in the case I.}%
\label{n3}%
\end{center}
\end{figure}
\begin{figure}
[h]
\begin{center}
\includegraphics[
height=2.6792in,
width=3.5786in
]%
{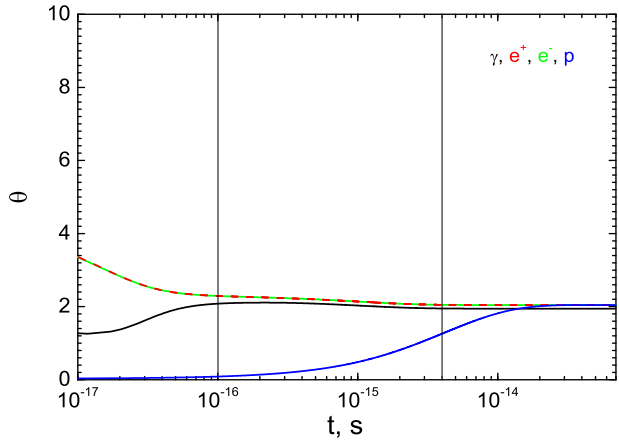}%
\caption{Depencence on time of dimensionless temperature for initial
conditions III. Colors are as in the case I. Pairs and photons acquire the
temperature at $t_{\mathrm{k}}\simeq10^{-16}$ sec.}%
\label{theta3}%
\end{center}
\end{figure}
\begin{figure}
[h]
\begin{center}
\includegraphics[
height=2.6481in,
width=3.6201in
]%
{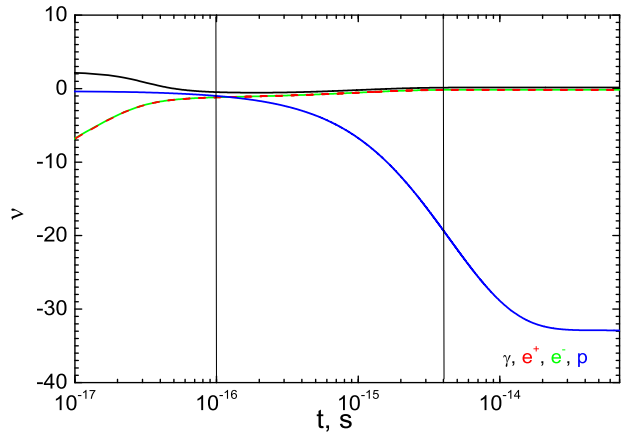}%
\caption{Depencence on time of dimensionless chemical potential for initial
conditions III. Colors are as in the case I. The chemical potentials equalize
at $t_{\mathrm{k}}\simeq10^{-16}$ sec.}%
\label{phi3}%
\end{center}
\end{figure}
\begin{figure}
[h]
\begin{center}
\includegraphics[
height=2.6048in,
width=3.5466in
]%
{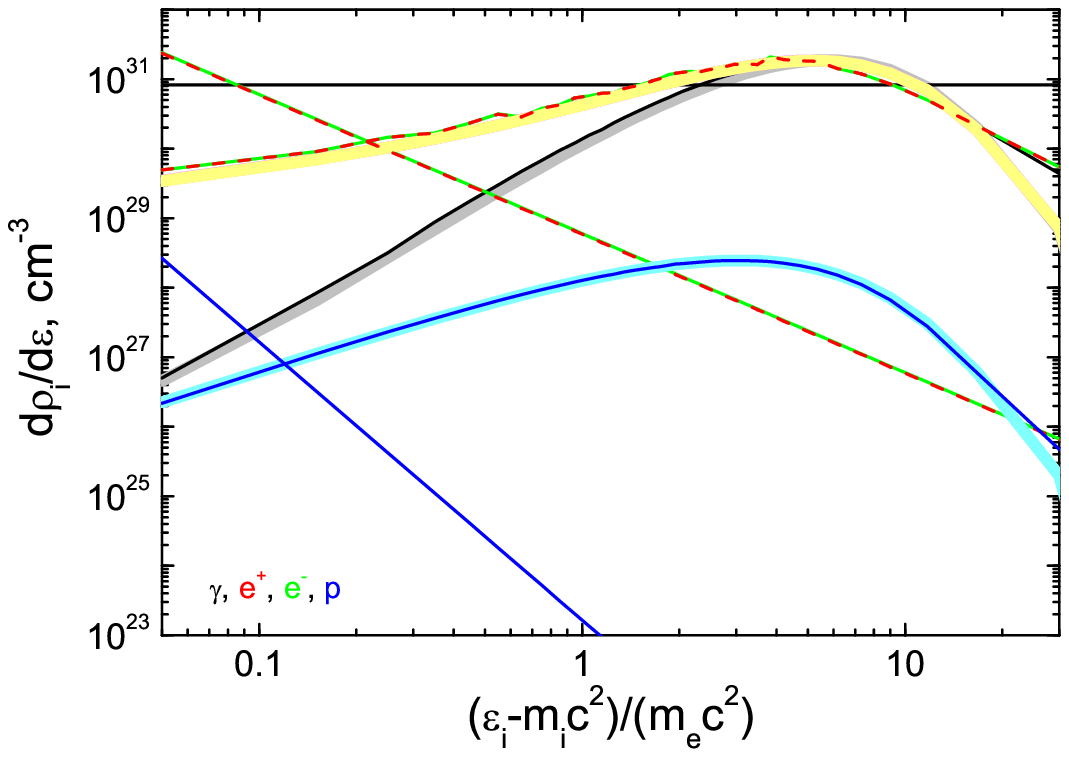}%
\caption{Initial and final spectral density as function of particle energy for
initial conditions III. The spectrum of protons is chosen to be steeper than
the one of electrons and positrons. Fits of the final spectra with chemical
potentials and temperatures are also shown.}%
\label{spectra3a}%
\end{center}
\end{figure}
The initial conditions are chosen in a way to get larger temperature in
thermal equilibrium, than in previous cases. Unlike the case II, the spectrum
of protons is chosen steeper than the spectrum of pairs in order to make them
colder in kinetic equilibrium.

Equipartition of energies between pairs and photons occurs earlier than in the
case I, at around $10^{-17}$\ sec, see fig. \ref{rho3}, since now
concentrations of particles are higher. Concentrations of pairs and photons
equalize at $3\times10^{-17}$\ sec, see fig. \ref{n3}. As in the case I, from
this moment temperatures and chemical potentials of electrons, positrons and
photons tend to be equal, see fig. \ref{theta3} and fig. \ref{phi3}
respectively, leading to kinetic equilibrium at around $t_{\mathrm{k}}%
\simeq10^{-16}$\ sec.

At the moment $t_{\mathrm{k}}$, shown by the vertical line on the left in fig.
\ref{theta3} and fig. \ref{phi3}, the temperature of photons and pairs is
$\theta_{\mathrm{k}}\simeq2.2$, the chemical potentials of these particles are
$\nu_{\mathrm{k}}\simeq-1.1$, while the temperature of protons, having well
established spectrum by this time, is just $\theta_{p}\simeq0.09$.

Thermal equilibrium is reached in the electron-positron-photon plasma at
around $t_{\mathrm{th}}\simeq4\times10^{-15}$\ sec, shown by the vertical line
on the right of fig. \ref{theta3} and fig. \ref{phi3}. Only at $4\times
10^{-14}$\ sec the temperature becomes common also with protons which are
heated up during this time. The temperature at this final stage is
$\theta_{\mathrm{th}}\simeq2$ while the chemical potential of protons is
$\nu_{p}\simeq-33$. Initial as well as final spectra are shown in fig.
\ref{spectra3a}.%

\begin{table}[tbp] \centering
\begin{tabular}
[c]{|c|c|c|c|}\hline\hline
& I & II & III\\\hline\hline
$\tau_{\mathrm{ch.eq}}^{\pm,\gamma}$, sec & $2.2\times10^{-13}$ &
$1.8\times10^{-14}$ & $9.5\times10^{-16}$\\\hline
$\tau_{\mathrm{ch.eq}}^{p}$, sec & $6\times10^{-13}$ & $1.8\times10^{-14}$ &
$5.5\times10^{-15}$\\\hline\hline
\end{tabular}
\caption{Relaxation time constant for cases I-III.}\label{tab3}%
\end{table}%
Since chemical potentials and temperatures approach their values in thermal
equilibrium exponentially, i.e. $\sim\exp(-t/\tau_{\mathrm{ch.eq}})$, we
determined the relaxation time constant $\tau_{\mathrm{ch.eq}}$ for each of
the cases considered from%
\begin{equation}
\tau_{\mathrm{ch.eq}}=\lim_{t\rightarrow\infty}\left[  \left(  F(t)-F(\infty
)\right)  \left(  \frac{dF}{dt}\right)  ^{-1}\right]  ,
\end{equation}
where the fugacity for a given sort of particle is given by (\ref{fugacity}).
Our results are presented in Tab. \ref{tab3}.

\section{Discussion and conclusions}

Results presented above clearly show the existence of two types of
equilibrium:\ the kinetic and the thermal ones. Kinetic equilibrium in
pair-photon plasma occurs when Ehlers \cite{1973rela.conf....1E} balance
conditions (\ref{Eb1}),(\ref{Eb2}) and (\ref{Eb3}) are satisfied so that
pair-creation, Compton and Bhabha/M{\o }ller scattering processes all come in
detailed balance. The electron-positron-photon plasma then is described by
common temperature and nonzero chemical potentials which are given by
(\ref{sumenergy}),(\ref{sumconc}) and (\ref{electronsnu}),(\ref{photonsnu}).
Protons at this stage may or may not have yet established equilibrium with the
spectrum (\ref{dk}), depending on the value of the baryon loading parameter
$\mathbf{B}$. When $\mathbf{B}$ is small, as in the case I, proton-proton
collision are inefficient since the rate (\ref{ppR}) is much smaller than
(\ref{epR}), and the proton spectrum is shaped up by the
proton-electron/positron collisions, reaching equilibrium form at a timescale
given by (\ref{epR}), when other particles are already in thermal equilibrium.
When $\mathbf{B}$ is large, as in the case II, protons have established their
equilibrium temperature at a timescale given by (\ref{ppR}), prior to the
moment when kinetic equilibrium in the pair-photon plasma is established.

As we have seen, the final spectra are completely insensitive to the initial
spectra, chosen to be flat as in the case I, power-law as in the case II, or
thermal ones.

The meaning of non-zero chemical potentials in kinetic equilibrium can be
understood as follows. The existence of a non-null chemical potential for
photons indicates the departure of the distribution function from the one
corresponding to the thermal equilibrium. Negative value of the chemical
potential generates an increase of the number of particles in order to
approach the one corresponding to the thermal equilibrium state. Positive
value of the chemical potential leads to the opposite effect, decreasing the
number of particles. Then, since the total number of particles increases (or
decreases), the energy is shared between larger (or smaller) number of
particles and the temperature decreases (or increases). Clearly, as thermal
equilibrium is approached, the chemical potential of photons tends to zero,
while the chemical potentials of electrons and positrons are given by
(\ref{thermalnu}), to guarantee an overall charge neutrality.

One of the basic assumptions in this work is that triple interactions are
slower than binary ones, allowing to use reaction rates for triple
interactions in kinetic equilibrium, explicitly depending solely on
temperature, chemical potentials and concentrations of particles. For pure
electron-positron plasma in the range of energies of interest (\ref{avenergy})
there is a hierarchy of relevant timescales: binary interactions are clearly
faster than triple ones. However, when protons are also present, the
proton-proton timescale may be shorter or longer than the corresponding binary
interactions timescales for the pure pair plasma. This violates our assumption
and therefore leads to loss of quantitative accuracy, although still keeping
qualitative results valid. In order to overcome this difficulty and produce
quantitatively precise results exact QED matrix elements must be used for
calculation of emission and absorption coefficients.

Notice that there is some discrepancy between our final spectra and their
thermal fits for high energy. This is due to poor energy resolution with
adopted grid. The result converges with higher resolutions, but it is limited
by the available computer memory. In addition, the code is quite
time-consuming and processor time increases with number of operation as third
power of the number of energy intervals.

In order to resolve proton-electron/positron scattering the number of energy
intervals should be increased as $M/m$ comparing to the case of pure pair
plasma. Even using inhomogeneous energy grid with uniform energy step up to
the peak of the spectrum $d\rho/d\varepsilon$\ and decreasing energy step as
$\varepsilon^{-1}$ for higher energies, we have obtained acceptable results
with about $10^{3}$ intervals for this reaction. Using such fine grid is
impossible in practice. On the other hand, a small parameter $m/M$ expansion
can be adopted. In this way we have introduced the mass scaling, described in
Appendix \ref{massscaling}, which gives quite good accuracy for about $10^{2}$
intervals in energy with inhomogeneous grid, described above. Finally, it is
important to stress that our code allows solution of the Boltzmann equations
for long time intervals and timescales, which may differ up to 10 orders of
magnitude, from electron-positron creation and annihilation process up to
proton-electron/positron scattering, see fig. \ref{n1}, unlike approaches
based on Monte-Carlo technique \cite{1997ApJ...486..903P}. This gives us the
possibility to follow thermalization process up to reaching steady solution,
i.e. thermal equilibrium.

The assumption of the constancy of the energy density is only valid if the
following three conditions are satisfied:

\begin{itemize}
\item plasma is optically thick for photons. This leads to the constraint on
the spatial dimensions $R_{0}\gg\left(  n_{\mathrm{th}}\sigma_{T}\right)
^{-1}\sim10^{-5}$ cm.

\item neutrinos are not produced. This gives the constraint on the temperature
$\theta\ll7\times10^{2}$.

\item plasma does not expand. Given $t_{\mathrm{dyn}}=\left(  \frac{1}{R}%
\frac{dR}{dt}\right)  ^{-1}\gg t_{\mathrm{th}}$, this leads to $R_{0}%
\gg10^{-2}$ cm.
\end{itemize}

To summarize, we have considered the evolution of initially nonequilibrium
optically thick electron-positron-photon plasma with proton loading up to
reaching thermal equilibrium on a timescale $t_{\mathrm{th}}\lesssim10^{-11} $
sec. Starting from arbitrary initial conditions we obtain kinetic equilibrium,
on a timescale $t_{\mathrm{k}}\lesssim10^{-14}$ sec, from first principles,
solving numerically the relativistic Boltzmann equation with collisional
integrals computed from exact QED matrix elements.

The general theoretical framework here presented can be further applied by
considering thermalization of different relativistic particles predicted by
extensions of the standard model of particle physics with the lepton plasma in
the early Universe. The occurence of thermalization process of
electron-positron plasma in GRBs on a much shorter timescale than the
characteristic acceleration time \cite{2000A&A...359..855R} is crucial. Such
acceleration timescales are indeed sharply bounded (shorter than $10^{3}$ sec
in the laboratory frame). Determination of thermalization timescales as
functions of the relevant parameters is important for high energy plasma
physics \cite{Ruffini2009},\cite{2008arXiv0801.0956T}. Finally, these results
can in principle be tested in laboratory experiments aiming the generation of
electron-positron pairs.

We thank the anonymous referee for comments which improved the presentation of
our results.%

\appendix

\section{Conservation laws}

\label{conslaws}

Conservation laws consist of baryon number, charge and energy conservations.
In addition, in binary reactions particle number is conserved.

Energy conservation law can be rewritten for the spectral density%
\begin{equation}
\frac{d}{dt}\sum_{i}\rho_{i}=0,\quad\mathrm{or}\quad\frac{d}{dt}\sum
_{i,\omega}Y_{i,\omega}=0, \label{energycons}%
\end{equation}
where%
\begin{equation}
Y_{i,\omega}=\int_{\epsilon_{i,\omega}-\Delta\epsilon_{i,\omega}/2}%
^{\epsilon_{i,\omega}+\Delta\epsilon_{i,\omega}/2}E_{i}d\epsilon.
\end{equation}
Particle's conservation law in binary reactions gives%
\begin{equation}
\frac{d}{dt}\sum_{i}n_{i}=0,\quad\mathrm{or}\quad\frac{d}{dt}\sum_{i,\omega
}\frac{Y_{i,\omega}}{\epsilon_{i,\omega}}=0. \label{numbercons}%
\end{equation}
Since baryonic number is conserved, therefore the number density of protons is
a constant%

\begin{subequations}
\begin{equation}
\frac{dn_{p}}{dt}=0. \label{baryoncons}%
\end{equation}
For electrically neutral plasma considered in this paper charge conservation
implies (\ref{chargecons}).

\section{Determination of temperature and chemical potentials in kinetic
equilibrium}

Consider distribution functions for photons and pairs in the most general form
(\ref{dk}). If one supposes that reaction rate for the Bhabha scattering
vanishes, i.e. there is equilibrium with respect to reaction
\end{subequations}
\begin{equation}
e^{+}+e^{-}\leftrightarrow+e^{+}{^{\prime}}+e^{-\prime},
\end{equation}
then the corresponding condition can be written in the following way
\begin{equation}
f_{+}(1-f_{+}{^{\prime}})f_{-}(1-f_{-}{^{\prime}})=f_{+}{^{\prime}}%
(1-f_{+})f_{-}{^{\prime}}(1+f_{-}), \label{Eb1}%
\end{equation}
where Bose-Einstein enhancement along with Pauli blocking factors are taken
into account for generality, it can be shown that electrons and positrons have
the same temperature%
\begin{equation}
\theta_{+}=\theta_{-}\equiv\theta_{\pm}, \label{kineticTe}%
\end{equation}
and they have arbitrary chemical potentials.

With (\ref{kineticTe}) analogous consideration for the Compton scattering
\begin{equation}
e^{\pm}+\gamma\leftrightarrow+e^{\pm}{^{\prime}}+\gamma^{\prime},
\end{equation}
gives%
\begin{equation}
f_{\pm}(1-f_{\pm}{^{\prime}})f_{\gamma}(1+f_{\gamma}{^{\prime}})=f_{\pm
}{^{\prime}}(1-f_{\pm})f_{\gamma}{^{\prime}}(1+f_{\gamma}), \label{Eb2}%
\end{equation}
and leads to equality of temperatures of pairs and photons
\begin{equation}
\theta_{\pm}=\theta_{\gamma}\equiv\theta_{k}, \label{kineticT}%
\end{equation}
with arbitrary chemical potentials. If, in addition, reaction rate in the
pair-creation and annihilation process%
\begin{equation}
e^{\pm}+e^{\mp}\leftrightarrow\gamma+\gamma^{\prime}%
\end{equation}
vanishes too, i.e. there is equilibrium with respect to pair production and
annihilation, with the corresponding condition,
\begin{equation}
f_{+}f_{-}(1+f_{\gamma})(1+f_{\gamma}{^{\prime}})=f_{\gamma}f_{\gamma
}{^{\prime}}(1-f_{+})(1-f_{-}), \label{Eb3}%
\end{equation}
it turns out that also chemical potentials of pairs and photons satisfy the
following condition%
\begin{equation}
\nu_{+}+\nu_{-}=2\nu_{\gamma}. \label{chempotcond}%
\end{equation}
However, since, generally speaking, $\nu_{\gamma}\neq0$ the condition
(\ref{chempotcond}) does not imply $\nu_{+}=\nu_{-}$. These considerations
were for the first time applied by Ehlers in \cite{1973rela.conf....1E}, see
also \cite{1963AcPP...23..629C}, and we will call (\ref{Eb1}),(\ref{Eb2}) and
(\ref{Eb3}) the \emph{Ehlers balance conditions}.

Analogous consideration for the detailed balance conditions in different
reactions lead to relations between temperatures and chemical potentials
summarized in table \ref{tab2}.%

\begin{table}[tbp] \centering
\begin{tabular}
[c]{|c|c|c|}\hline\hline
& Interaction & Parameters of DFs\\\hline\hline
I & $e^{+}e^{-}$ scattering & $\theta_{+}=\theta_{-}$, $\forall\nu_{+}$%
,$\nu_{-}$\\\hline
II & $e^{\pm}p$ scattering & $\theta_{p}=\theta_{\pm}$, $\forall\nu_{\pm}%
$,$\nu_{p}$\\\hline
III & $e^{\pm}\gamma$ scattering & $\theta_{\gamma}=\theta_{\pm}$, $\forall
\nu_{\gamma}$,$\nu_{\pm}$\\\hline
IV & pair production & $\nu_{+}+\nu_{-}=2\nu_{\gamma}$, if $\theta_{\gamma
}=\theta_{\pm}$\\\hline
V & Tripe interactions & $\nu_{\gamma}$, $\nu_{\pm}=0$, if $\theta_{\gamma
}=\theta_{\pm}$\\\hline\hline
\end{tabular}
\caption{Relations between parameters of equilibrium DFs fulfilling detailed
balance conditions for the reactions shown in Tab.~\ref{tab1}.}\label{tab2}
\end{table}%

The timescales of pair production and annihilation processes as well as
Compton scattering are nearly equal in the range of energies of interest and
are given by (\ref{ttau}). Therefore, kinetic equilibrium\ is first
established simultaneously for electrons, positrons and photons. They reach
the same temperature, but with chemical potentials different from zero. Later
on, the temperatures of this electron-positron-photon plasma and the one of
protons reach a common value.

In order to find temperatures and chemical potentials we have to implement the
following constraints:\ energy conservation (\ref{energycons}), particle
number conservation (\ref{numbercons}), charge conservation (\ref{chargecons}%
), condition for the chemical potentials (\ref{chempotcond}).

Given (\ref{dk}) we have for photons%
\begin{equation}
\frac{\rho_{\gamma}}{n_{\gamma}mc^{2}}=3\theta_{\gamma},\quad n_{\gamma}%
=\frac{1}{V_{0}}\exp\left(  \frac{\nu_{\gamma}}{\theta_{\gamma}}\right)
2\theta_{\gamma}^{3}, \label{kineqphotons}%
\end{equation}
for pairs%
\begin{equation}
\frac{\rho_{\pm}}{n_{\pm}mc^{2}}=j_{2}(\theta_{\pm}),\quad n_{\pm}=\frac
{1}{V_{0}}\exp\left(  \frac{\nu_{\pm}}{\theta_{\pm}}\right)  j_{1}(\theta
_{\pm}), \label{kineqpairs}%
\end{equation}
and for protons%
\begin{gather}
\frac{\rho_{p}}{Mn_{p}c^{2}}=1+\frac{3}{2}\frac{m}{M}\theta_{p}%
,\label{kineqprotons1}\\
n_{p}=\frac{1}{V_{0}}\sqrt{\frac{\pi}{2}}\left(  \frac{M}{m}\right)
^{3/2}\exp\left(  \frac{\nu_{p}-\frac{M}{m}}{\theta_{p}}\right)  \theta
_{p}^{\frac{3}{2}}, \label{kineqprotons2}%
\end{gather}
where we assumed that protons are nonrelativistic, and we denoted the Compton
volume%
\begin{equation}
V_{0}=\frac{1}{8\pi}\left(  \frac{2\pi\hbar}{mc}\right)  ^{3},
\end{equation}
and functions $j_{1}$ and $j_{2}$\ are defined as%
\begin{align}
j_{1}(\theta)  &  =\theta K_{2}(\theta^{-1})\rightarrow\left\{
\begin{array}
[c]{cc}%
\sqrt{\frac{\pi}{2}}e^{-\frac{1}{\theta}}\theta^{3/2}, & \theta\rightarrow0\\
2\theta^{3}, & \theta\rightarrow\infty
\end{array}
\right.  ,\\
j_{2}(\theta)  &  =\frac{3K_{3}(\theta^{-1})+K_{1}(\theta^{-1})}{4K_{2}%
(\theta^{-1})}\rightarrow\left\{
\begin{array}
[c]{cc}%
1+\frac{3\theta}{2}, & \theta\rightarrow0\\
3\theta, & \theta\rightarrow\infty
\end{array}
\right.  .
\end{align}

For pure electron-positron-photon plasma in kinetic equilibrium, summing up
energy densities in (\ref{kineqphotons}),(\ref{kineqpairs}) and using
(\ref{kineticTe}),(\ref{kineticT}) and (\ref{chempotcond}) we obtain%
\begin{equation}
\sum_{e^{+},e^{-},\gamma}\rho_{i}=\frac{2mc^{2}}{V_{0}}\exp\left(  \frac
{\nu_{\mathrm{k}}}{\theta_{\mathrm{k}}}\right)  \left[  3\theta^{4}%
+j_{1}(\theta_{k})j_{2}(\theta_{\mathrm{k}})\right]  , \label{energypair}%
\end{equation}
and analogously for number densities we get%
\begin{equation}
\sum_{e^{+},e^{-},\gamma}n_{i}=\frac{2}{V_{0}}\exp\left(  \frac{\nu
_{\mathrm{k}}}{\theta_{\mathrm{k}}}\right)  \left[  \theta_{\mathrm{k}}%
^{3}+j_{1}(\theta_{\mathrm{k}})\right]  . \label{numberpair}%
\end{equation}
From (\ref{energypair}) and (\ref{numberpair}) two unknowns, $\nu_{k}$ and
$\theta_{\mathrm{k}}$ can be found.

When protons are present, in most cases the electron-positron-photon plasma
reaches kinetic equilibrium first, while protons join the plasma later. In
that case, the temperature of protons $\theta_{p}$ is different from the rest
of particles, so while $\theta_{+}=\theta_{-}=\theta_{\gamma}=\theta
_{\mathrm{k}}$, $\theta_{p}\neq\theta_{\mathrm{k}}$.

Then summing up energy densities in (\ref{kineqphotons}),(\ref{kineqpairs}) we
obtain%
\begin{gather}
\sum_{e^{+},e^{-},\gamma}\rho_{i}=\frac{mc^{2}}{V_{0}}\left\{  \left[
1-\frac{n_{p}V_{0}}{j_{1}(\theta_{\mathrm{k}})}\exp\left(  -\frac{\nu_{+}%
}{\theta_{\mathrm{k}}}\right)  \right]  ^{\frac{1}{2}}\times\right.
\label{sumenergy}\\
\left.  \times6\theta_{\mathrm{k}}^{4}\exp\left(  \frac{\nu_{+}}%
{\theta_{\mathrm{k}}}\right)  +\left[  2j_{1}(\theta_{k})\exp\left(  \frac
{\nu_{+}}{\theta_{\mathrm{k}}}\right)  -n_{p}V_{0}\right]  j_{2}%
(\theta_{\mathrm{k}})\right\}  ,\nonumber
\end{gather}
and analogously for number densities we get%
\begin{gather}
\sum_{e^{+},e^{-},\gamma}n_{i}=\frac{1}{V_{0}}\left\{  \left[  1-\frac
{n_{p}V_{0}}{j_{1}(\theta_{\mathrm{k}})}\exp\left(  -\frac{\nu_{+}}%
{\theta_{\mathrm{k}}}\right)  \right]  ^{\frac{1}{2}}\times\right.
\label{sumconc}\\
\left.  \times6\theta_{\mathrm{k}}^{4}\exp\left(  \frac{\nu_{+}}%
{\theta_{\mathrm{k}}}\right)  +2j_{1}(\theta_{\mathrm{k}})\exp\left(
\frac{\nu_{+}}{\theta_{\mathrm{k}}}\right)  \right\}  .\nonumber
\end{gather}

From (\ref{sumenergy}) and (\ref{sumconc}) two unknowns, $\nu_{+}$ and
$\theta_{\mathrm{k}}$ can be found. Then the rest of the chemical potentials
are obtained from%
\begin{gather}
\exp\left(  \frac{\nu_{-}}{\theta_{\mathrm{k}}}\right)  =\exp\left(  \frac
{\nu_{+}}{\theta_{\mathrm{k}}}\right)  +\frac{n_{p}V_{0}}{j_{1}(\theta
_{\mathrm{k}})},\label{electronsnu}\\
\exp\left(  \frac{\nu_{\gamma}}{\theta_{\mathrm{k}}}\right)  =\exp\left(
\frac{\nu_{+}}{\theta_{\mathrm{k}}}\right)  \left[  1+\frac{n_{p}V_{0}}%
{j_{1}(\theta_{\mathrm{k}})}\exp\left(  -\frac{\nu_{+}}{\theta_{\mathrm{k}}%
}\right)  \right]  ^{\frac{1}{2}}, \label{photonsnu}%
\end{gather}
The temperature and chemical potential of protons can be found separately from
(\ref{kineqprotons1}),(\ref{kineqprotons2}).

In thermal equilibrium $\nu_{\gamma}$ vanishes and one has%
\begin{equation}
\nu_{-}=\theta_{\mathrm{k}}\operatorname{arcsinh}\left[  \frac{n_{p}V_{0}%
}{2j_{1}(\theta_{\mathrm{k}})}\right]  ,\qquad\nu_{+}=-\nu_{-},
\label{thermalnu}%
\end{equation}
which both reduce to $\nu_{-}=\nu_{+}=0$ for $n_{p}=0$. At the same time, for
$n_{p}>0$ one always has $\nu_{-}>0$ and $\nu_{+}<0$ in thermal equilibrium.
The chemical potential of protons in thermal equilibrium is determined from
(\ref{kineqprotons2}) for $\theta_{\mathrm{k}}=\theta_{\mathrm{th}}$, where
$\theta_{\mathrm{th}}$ is the temperature in thermal equilibrium.%

\begin{widetext}%

\section{Binary interactions}

\label{2body}

\subsection{Compton scattering $\gamma e^{\pm}\rightarrow\gamma^{\prime}%
e^{\pm\prime}$}

\label{compton}

The time evolution of the distribution functions of photons and pair particles
due to Compton scattering may be described by \cite{1981els..book.....L}%
,\cite{1979MAt..book.....O}%
\begin{equation}
\left(  \frac{\partial f_{\gamma}(\mathbf{k},t)}{\partial t}\right)  _{\gamma
e^{\pm}\rightarrow\gamma^{\prime}e^{\pm\prime}}=\int d\mathbf{k}^{\prime
}d\mathbf{p}d\mathbf{p}^{\prime}Vw_{\mathbf{k}^{\prime},\mathbf{p}^{\prime
};\mathbf{k},\mathbf{p}}[f_{\gamma}({\mathbf{k}^{\prime},t)}f_{\pm
}({\mathbf{p}^{\prime}},t)-f_{\gamma}(\mathbf{k},t)f_{\pm}(\mathbf{p},t)],
\label{Comptongamma}%
\end{equation}%
\begin{equation}
\left(  \frac{\partial f_{\pm}(\mathbf{p},t)}{\partial t}\right)  _{\gamma
e^{\pm}\rightarrow\gamma^{\prime}e^{\pm\prime}}=\int d\mathbf{k}%
d\mathbf{k}^{\prime}d\mathbf{p}^{\prime}Vw_{\mathbf{k}^{\prime},\mathbf{p}%
^{\prime};\mathbf{k},\mathbf{p}}[f_{\gamma}({\mathbf{k}^{\prime}},t)f_{\pm
}({\mathbf{p}^{\prime}},t)-f_{\gamma}(\mathbf{k},t)f_{\pm}(\mathbf{p},t)],
\label{Comptone}%
\end{equation}
where
\begin{equation}
w_{\mathbf{k}^{\prime},\mathbf{p}^{\prime};\mathbf{k},\mathbf{p}}=\frac
{\hbar^{2}c^{6}}{(2\pi)^{2}V}\delta(\epsilon_{\gamma}-\epsilon_{\pm}%
-\epsilon_{\gamma}^{\prime}-\epsilon_{\pm}^{\prime})\delta(\mathbf{k}%
+\mathbf{p}-\mathbf{k}^{\prime}-\mathbf{p}^{\prime})\frac{|M_{fi}|^{2}%
}{16\epsilon_{\gamma}\epsilon_{\pm}\epsilon_{\gamma}^{\prime}\epsilon_{\pm
}^{\prime}}, \label{df-abs}%
\end{equation}
is the probability of the process,%

\begin{align}
|M_{fi}|^{2}  &  =2^{6}\pi^{2}\alpha^{2}\left[  \frac{m^{2}c^{2}}{s-m^{2}%
c^{2}}+\frac{m^{2}c^{2}}{u-m^{2}c^{2}}+\left(  \frac{m^{2}c^{2}}{s-m^{2}c^{2}%
}+\frac{m^{2}c^{2}}{u-m^{2}c^{2}}\right)  ^{2}\right. \nonumber\\
&  \left.  -\frac{1}{4}\left(  \frac{s-m^{2}c^{2}}{u-m^{2}c^{2}}+\frac
{u-m^{2}c^{2}}{s-m^{2}c^{2}}\right)  \right]  , \label{M_fi_gamma1}%
\end{align}
is the square of the matrix element, $s=(\mathfrak{p}+\mathfrak{k})^{2}$ and
$u=(\mathfrak{p}-\mathfrak{k}^{\prime})^{2}$ are invariants, $\mathfrak{k}%
=(\epsilon_{\gamma}/c)(1,\mathbf{e}_{\gamma})$ and $\mathfrak{p}%
=(\epsilon_{\pm}/c)(1,\beta_{\pm}\mathbf{e}_{\pm})$ are energy-momentum four
vectors of photons and electrons, respectively, $d\mathbf{p}=d\epsilon_{\pm
}do\epsilon_{\pm}^{2}\beta_{\pm}/c^{3}$, $d\mathbf{k}^{\prime}=d\epsilon
_{\gamma}^{\prime}\epsilon_{\gamma}^{\prime2}do_{\gamma}^{\prime}/c^{3}$ and
$do=d\mu d\phi$.

The energies of photon and positron (electron) after the scattering are%
\begin{equation}
\epsilon_{\gamma}^{\prime}=\frac{\epsilon_{\pm}\epsilon_{\gamma}(1-\beta_{\pm
}\mathbf{b}_{\pm}\mathbf{\cdot}\mathbf{b}_{\gamma})}{\epsilon_{\pm}%
(1-\beta_{\pm}\mathbf{b}_{\pm}\mathbf{\cdot}\mathbf{b}_{\gamma}^{\prime
})+\epsilon_{\gamma}(1-\mathbf{b}_{\gamma}\mathbf{\cdot}\mathbf{b}_{\gamma
}^{\prime})}\,,\qquad\,\epsilon_{\pm}^{\prime}=\epsilon_{\pm}+\epsilon
_{\gamma}-\epsilon_{\gamma}^{\prime}\,,
\end{equation}
$\mathbf{b}_{i}=\mathbf{p}_{i}/p$, $\mathbf{b}_{i}^{\prime}=\mathbf{p}%
_{i}^{\prime}/p^{\prime}$, $\mathbf{b}_{\pm}^{\prime}=(\beta_{\pm}%
\epsilon_{\pm}\mathbf{b}_{\pm}+\epsilon_{\gamma}\mathbf{b}_{\gamma}%
-\epsilon_{\gamma}^{\prime}\mathbf{b}_{\gamma}^{\prime})/(\beta_{\pm}^{\prime
}\epsilon_{\pm}^{\prime})$.

For photons, the absorption coefficient (\ref{comptonabscoef}) in the
Boltzmann equations (\ref{BE}) is
\begin{equation}
\chi_{\gamma}^{\gamma e^{\pm}\rightarrow\gamma^{\prime}e^{\pm\prime}}%
f_{\gamma}=-\frac{1}{c}\left(  \frac{\partial f_{\gamma}}{\partial t}\right)
_{\gamma e^{\pm}\rightarrow\gamma^{\prime}e^{\pm\prime}}^{\mathrm{abs}}=\int
dn_{\pm}do_{\gamma}^{\prime}J_{\mathrm{cs}}\frac{\epsilon_{\gamma}^{\prime
}|M_{fi}|^{2}\hbar^{2}c^{2}}{16\epsilon_{\pm}\epsilon_{\gamma}\epsilon_{\pm
}^{\prime}}f_{\gamma}, \label{chi-fgamma}%
\end{equation}
where $dn_{i}=d\epsilon_{i}do_{i}\epsilon_{i}^{2}\beta_{i}f_{i}/c^{3}%
=d\epsilon_{i}do_{i}E_{i}/(2\pi\epsilon_{i})$.

From equations (\ref{Comptongamma}) and (\ref{chi-fgamma}), we can write the
absorption coefficient for photon energy density $E_{\gamma}$ averaged over
the $\epsilon,\mu$-grid with zone numbers $\omega$ and $k$ as
\begin{align}
(\chi E)_{\gamma,\omega}^{\gamma e^{\pm}\rightarrow\gamma^{\prime}e^{\pm
\prime}}  &  \equiv\frac{1}{\Delta\epsilon_{\gamma,\omega}}\int_{\epsilon
_{\gamma}\in\Delta\epsilon_{\gamma,\omega}}d\epsilon_{\gamma}d\mu_{\gamma
}(\chi E)_{\gamma}^{\gamma e^{\pm}\rightarrow\gamma^{\prime}e^{\pm\prime}%
}=\nonumber\\
&  =\frac{1}{\Delta\epsilon_{\gamma,\omega}}\int_{\epsilon_{\gamma}\in
\Delta\epsilon_{\gamma,\omega}}dn_{\gamma}dn_{\pm}do_{\gamma}^{\prime
}J_{\mathrm{cs}}\frac{\epsilon_{\gamma}^{\prime}|M_{fi}|^{2}\hbar^{2}c^{2}%
}{16\epsilon_{\pm}\epsilon_{\pm}^{\prime}}, \label{chiE1}%
\end{align}
where the Jacobian of the transformation is%
\begin{equation}
J_{\mathrm{cs}}=\frac{\epsilon_{\gamma}^{\prime}\epsilon_{\pm}^{\prime}%
}{\epsilon_{\gamma}\epsilon_{\pm}\left(  1-\beta_{\pm}\mathbf{b}_{\gamma
}\mathbf{\cdot b}_{\pm}\right)  }.
\end{equation}
Similar integrations can be performed for the other terms of equations
(\ref{Comptongamma}), (\ref{Comptone}), and we obtain%
\begin{align}
\eta_{\gamma,\omega}^{\gamma e^{\pm}\rightarrow\gamma^{\prime}e^{\pm\prime}}
&  =\frac{1}{\Delta\epsilon_{\gamma,\omega}}\int_{\epsilon_{\gamma}^{\prime
}\in\Delta\epsilon_{\gamma,\omega}}dn_{\gamma}dn_{\pm}do_{\gamma}^{\prime
}J_{\mathrm{cs}}\frac{\epsilon_{\gamma}^{\prime2}|M_{fi}|^{2}\hbar^{2}c^{2}%
}{16\epsilon_{\pm}\epsilon_{\gamma}\epsilon_{\pm}^{\prime}},\label{chiE2}\\
\eta_{\pm,\omega}^{\gamma e^{\pm}\rightarrow\gamma^{\prime}e^{\pm\prime}}  &
=\frac{1}{\Delta\epsilon_{\pm,\omega}}\int_{\epsilon_{\pm}^{\prime}\in
\Delta\epsilon_{\pm,\omega}}dn_{\gamma}dn_{\pm}do_{\gamma}^{\prime
}J_{\mathrm{cs}}\frac{\epsilon_{\gamma}^{\prime}|M_{fi}|^{2}\hbar^{2}c^{2}%
}{16\epsilon_{\pm}\epsilon_{\gamma}},\label{chiE3}\\
(\chi E)_{\pm,\omega}^{\gamma e^{\pm}\rightarrow\gamma^{\prime}e^{\pm\prime}}
&  =\frac{1}{\Delta\epsilon_{\pm,\omega}}\int_{\epsilon_{\pm}\in\Delta
\epsilon_{\pm,\omega}}dn_{\gamma}dn_{\pm}do_{\gamma}^{\prime}J_{\mathrm{cs}%
}\frac{\epsilon_{\gamma}^{\prime}|M_{fi}|^{2}\hbar^{2}c^{2}}{16\epsilon
_{\gamma}\epsilon_{\pm}^{\prime}}. \label{chiE4}%
\end{align}

In order to perform integrals (\ref{chiE1})-(\ref{chiE4}) numerically over
$\phi$ ($0\leq\phi\leq2\pi$) we introduce a uniform grid $\phi_{l\mp1/2}$ with
$1\leq l\leq l_{\mathrm{max}}$ and $\Delta\phi_{l}=(\phi_{l+1/2}-\phi
_{l-1/2})/2=2\pi/l_{\mathrm{max}}$. We assume that any function of $\phi$ in
equations (\ref{chiE1})-(\ref{chiE2}) in the interval $\Delta\phi_{j}$ is
equal to its value at $\phi=\phi_{j}=(\phi_{l-1/2}+\phi_{l+1/2})/2$. It is
necessary to integrate over $\phi$ only once at the beginning of calculations.
The number of intervals of the $\phi$-grid depends on the average energy of
particles and is typically taken as $l_{\mathrm{max}}=2k_{\mathrm{max}}=64$.

\subsection{Pair creation and annihilation $\gamma_{1}\gamma_{2}%
\rightleftarrows e^{-}e^{+}$}

\label{pair}

The rates of change of the distribution function due to pair creation and
annihilation are%

\begin{equation}
\left(  \frac{\partial f_{\gamma_{j}}(\mathbf{k}_{i},t)}{\partial t}\right)
_{\gamma_{1}\gamma_{2}\rightarrow e^{-}e^{+}}=-\int d\mathbf{k}_{j}%
d\mathbf{p}_{-}d\mathbf{p}_{+}Vw_{\mathbf{p}_{-},\mathbf{p}_{+};\mathbf{k}%
_{1},\mathbf{k}_{2}}f_{\gamma_{1}}(\mathbf{k}_{1},t)f_{\gamma_{2}}%
(\mathbf{k}_{2},t)\,, \label{fgamma1}%
\end{equation}%
\begin{equation}
\left(  \frac{\partial f_{\gamma_{i}}(\mathbf{k}_{i},t)}{\partial t}\right)
_{e^{-}e^{+}\rightarrow\gamma_{1}\gamma_{2}}=\int d\mathbf{k}_{j}%
d\mathbf{p}_{-}d\mathbf{p}_{+}Vw_{\mathbf{k}_{1},\mathbf{k}_{2};\mathbf{p}%
_{-},\mathbf{p}_{+}}f_{-}(\mathbf{p}_{-},t)f_{+}(\mathbf{p}_{+},t)\,,
\end{equation}
for $i=1,~j=2$, and for $j=1,~i=2$.%

\begin{equation}
\left(  \frac{\partial f_{\pm}(\mathbf{p}_{\pm},t)}{\partial t}\right)
_{\gamma_{1}\gamma_{2}\rightarrow e^{-}e^{+}}=\int d\mathbf{p}_{\mp
}d\mathbf{k}_{1}d\mathbf{k}_{2}Vw_{\mathbf{p}_{-},\mathbf{p}_{+}%
;\mathbf{k}_{1},\mathbf{k}_{2}}f_{\gamma}(\mathbf{k}_{1},t)f_{\gamma
}(\mathbf{k}_{2},t)\,,
\end{equation}%
\begin{equation}
\left(  \frac{\partial f_{\pm}(\mathbf{p}_{\pm},t)}{\partial t}\right)
_{e^{-}e^{+}\rightarrow\gamma_{1}\gamma_{2}}=-\int d\mathbf{p}_{\mp
}d\mathbf{k}_{1}d\mathbf{k}_{2}Vw_{\mathbf{k}_{1},\mathbf{k}_{2}%
;\mathbf{p}_{-},\mathbf{p}_{+}}f_{-}(\mathbf{p}_{-},t)f_{+}(\mathbf{p}%
_{+},t)\,, \label{fe+}%
\end{equation}
where
\begin{equation}
w_{\mathbf{p}_{-},\mathbf{p}_{+};\mathbf{k}_{1},\mathbf{k}_{2}}=\frac
{\hbar^{2}c^{6}}{(2\pi)^{2}V}\delta(\epsilon_{-}+\epsilon_{+}-\epsilon
_{1}-\epsilon_{2})\delta(\mathbf{p}_{-}+\mathbf{p}_{+}-\mathbf{k}%
_{1}-\mathbf{k}_{2})\frac{|M_{fi}|^{2}}{16\epsilon_{-}\epsilon_{+}\epsilon
_{1}\epsilon_{2}}.
\end{equation}
Here, the matrix element $|M_{fi}|^{2}$ is given by equation
(\ref{M_fi_gamma1}) with the new invariants $s=(\mathfrak{p}_{-}%
-\mathfrak{k}_{1})^{2}$ and $u=(\mathfrak{p}_{-}-\mathfrak{k}_{2})^{2}$, see
\cite{1982els..book.....B}.

The energies of photons created via annihilation of a $e^{\pm}$ pair are
\begin{equation}
\epsilon_{1}(\mathbf{b}_{1})=\frac{m^{2}c^{4}+\epsilon_{-}\epsilon_{+}%
(1-\beta_{-}\beta_{+}\mathbf{b}_{-}\mathbf{\cdot}\mathbf{b}_{+})}{\epsilon
_{-}(1-\beta_{-}\mathbf{b}_{-}\mathbf{\cdot}\mathbf{b}_{1})+\epsilon
_{+}(1-\beta_{+}\mathbf{b}_{+}\mathbf{\cdot}\mathbf{b}_{1})}\,,\qquad
\epsilon_{2}(\mathbf{b}_{1})=\epsilon_{-}+\epsilon_{+}-\epsilon_{1}\,,
\end{equation}
while the energies of pair particles created by two photons are found from%
\begin{equation}
\epsilon_{-}(\mathbf{b}_{-})=\frac{B\mp\sqrt{B^{2}-AC}}{A}\,,\qquad
\epsilon_{+}(\mathbf{b}_{-})=\epsilon_{1}+\epsilon_{2}-\epsilon_{-}\,,
\label{A27}%
\end{equation}
where $A=(\epsilon_{1}+\epsilon_{2})^{2}-[(\epsilon_{1}\mathbf{b}_{1}%
+\epsilon_{2}\mathbf{b}_{2})\mathbf{\cdot}\mathbf{b}_{-}]^{2}$, $B=(\epsilon
_{1}+\epsilon_{2})\epsilon_{1}\epsilon_{2}(1-\mathbf{b}_{1}\mathbf{\cdot
}\mathbf{b}_{2})$, $C=m_{e}^{2}c^{4}[(\epsilon_{1}\mathbf{b}_{1}+\epsilon
_{2}\mathbf{b}_{2})\mathbf{\cdot}\mathbf{b}_{-}]^{2}+\epsilon_{1}^{2}%
\epsilon_{2}^{2}(1-\mathbf{b}_{1}\mathbf{\cdot}\mathbf{b}_{2})^{2}$. Only one
root in equation (\ref{A27}) has to be chosen. From energy-momentum
conservation
\begin{equation}
\mathfrak{k}_{1}+\mathfrak{k}_{2}-\mathfrak{p}_{-}=\mathfrak{p}_{+},
\end{equation}
taking square from the energy part we have%
\begin{equation}
\epsilon_{1}^{2}+\epsilon_{2}^{2}+\epsilon_{-}^{2}+2\epsilon_{1}\epsilon
_{2}-2\epsilon_{1}\epsilon_{-}-2\epsilon_{2}\epsilon_{-}=\epsilon_{+}^{2},
\end{equation}
and taking square from the momentum part we get%
\begin{equation}
\epsilon_{1}^{2}+\epsilon_{2}^{2}+\epsilon_{-}^{2}\beta_{-}^{2}+2\epsilon
_{1}\epsilon_{2}\mathbf{b}_{1}\mathbf{\cdot b}_{2}-2\epsilon_{1}\epsilon
_{-}\beta_{-}\mathbf{b}_{1}\mathbf{\cdot b}_{-}-2\epsilon_{2}\epsilon_{-}%
\beta_{-}\mathbf{b}_{2}\mathbf{\cdot b}_{-}=(\epsilon_{+}\beta_{+})^{2}.
\end{equation}
There are no additional roots because of the arbitrary $\mathbf{e}_{+}$%
\begin{gather}
\epsilon_{1}\epsilon_{2}(1-\mathbf{b}_{1}\mathbf{\cdot b}_{2})-\epsilon
_{1}\epsilon_{-}(1-\beta_{-}\mathbf{b}_{1}\mathbf{\cdot b}_{-})-\epsilon
_{2}\epsilon_{-}(1-\beta\mathbf{b}_{2}\mathbf{\cdot b}_{-})=0,\\
\epsilon_{-}\beta_{-}(\epsilon_{1}\mathbf{b}_{1}+\epsilon_{2}\mathbf{b}%
_{2})\mathbf{\cdot b}_{-}=\epsilon_{-}(\epsilon_{1}+\epsilon_{2})-\epsilon
_{1}\epsilon_{2}(1-\mathbf{b}_{1}\mathbf{\cdot b}_{2}).\nonumber
\end{gather}
Eliminating $\beta$ we obtain%
\begin{align}
&  \epsilon_{1}^{2}\epsilon_{2}^{2}(1-\mathbf{b}_{1}\mathbf{\cdot b}_{2}%
)^{2}-2\epsilon_{1}\epsilon_{2}(1-\mathbf{b}_{1}\mathbf{\cdot b}_{2}%
)(\epsilon_{1}+\epsilon_{2})\epsilon_{-}+\left\{  (\epsilon_{1}+\epsilon
_{2})^{2}-\left[  (\epsilon_{1}\mathbf{b}_{1}+\epsilon_{2}\mathbf{b}%
_{2})\mathbf{\cdot b}_{-}\right]  ^{2}\right\}  \epsilon_{-}^{2}=\nonumber\\
&  =\left[  (\epsilon_{1}\mathbf{b}_{1}+\epsilon_{2}\mathbf{b}_{2}%
)\mathbf{\cdot b}_{-}\right]  (-m^{2}),
\end{align}
Therefore, the condition to be checked reads%
\begin{equation}
\epsilon_{-}\beta_{-}\left[  (\epsilon_{1}\mathbf{b}_{1}+\epsilon
_{2}\mathbf{b}_{2})\mathbf{\cdot b}_{-}\right]  ^{2}=\left[  \epsilon
_{-}(\epsilon_{1}+\epsilon_{2})-(\epsilon_{1}\epsilon_{2})(1-\mathbf{b}%
_{1}\mathbf{\cdot b}_{2})\right]  \left[  (\epsilon_{1}\mathbf{b}_{1}%
+\epsilon_{2}\mathbf{b}_{2})\mathbf{\cdot b}_{-}\right]  \geq0.
\end{equation}

Finally, integration of equations (\ref{fgamma1})-(\ref{fe+}) yields%
\begin{align}
\eta_{\gamma,\omega}^{e^{-}e^{+}\rightarrow\gamma_{1}\gamma_{2}}  &  =\frac
{1}{\Delta\epsilon_{\gamma,\omega}}\left(  \int_{\epsilon_{1}\in\Delta
\epsilon_{\gamma,\omega}}d^{2}n_{\pm}J_{\mathrm{ca}}\frac{\epsilon_{1}%
^{2}|M_{fi}|^{2}\hbar^{2}c^{2}}{16\epsilon_{-}\epsilon_{+}\epsilon_{2}}%
+\int_{\epsilon_{2}\in\Delta\epsilon_{\gamma,\omega}}d^{2}n_{\pm
}J_{\mathrm{ca}}\frac{\epsilon_{1}|M_{fi}|^{2}\hbar^{2}c^{2}}{16\epsilon
_{-}\epsilon_{+}}\right)  ,\label{A28}\\
(\chi E)_{e,\omega}^{e^{-}e^{+}\rightarrow\gamma_{1}\gamma_{2}}  &  =\frac
{1}{\Delta\epsilon_{e,\omega}}\left(  \int_{\epsilon_{-}\in\Delta
\epsilon_{e,\omega}}d^{2}n_{\pm}J_{\mathrm{ca}}\frac{\epsilon_{1}|M_{fi}%
|^{2}\hbar^{2}c^{2}}{16\epsilon_{+}\epsilon_{2}}+\int_{\epsilon_{+}\in
\Delta\epsilon_{e,\omega}}d^{2}n_{\pm}J_{\mathrm{ca}}\frac{\epsilon_{1}%
|M_{fi}|^{2}\hbar^{2}c^{2}}{16\epsilon_{-}\epsilon_{2}}\right)  ,\label{A29}\\
(\chi E)_{\gamma,\omega}^{\gamma_{1}\gamma_{2}\rightarrow e^{-}e^{+}}  &
=\frac{1}{\Delta\epsilon_{\gamma,\omega}}\left(  \int_{\epsilon_{1}\in
\Delta\epsilon_{\gamma,\omega}}d^{2}n_{\gamma}J_{\mathrm{ca}}\frac
{\epsilon_{-}\beta_{-}|M_{fi}|^{2}\hbar^{2}c^{2}}{16\epsilon_{2}\epsilon_{+}%
}+\int_{\epsilon_{2}\in\Delta\epsilon_{\gamma,\omega}}d^{2}n_{\gamma
}J_{\mathrm{ca}}\frac{\epsilon_{-}\beta_{-}|M_{fi}|^{2}\hbar^{2}c^{2}%
}{16\epsilon_{1}\epsilon_{+}}\right)  ,\label{A30}\\
\eta_{e,\omega}^{\gamma_{1}\gamma_{2}\rightarrow e^{-}e^{+}}  &  =\frac
{1}{\Delta\epsilon_{e,\omega}}\left(  \int_{\epsilon_{-}\in\Delta
\epsilon_{e,\omega}}d^{2}n_{\gamma}J_{\mathrm{ca}}\frac{\epsilon_{-}^{2}%
\beta_{-}|M_{fi}|^{2}\hbar^{2}c^{2}}{16\epsilon_{1}\epsilon_{2}\epsilon_{+}%
}+\int_{\epsilon_{+}\in\Delta\epsilon_{e,\omega}}d^{2}n_{\gamma}%
J_{\mathrm{ca}}\frac{\epsilon_{-}\beta_{-}|M_{fi}|^{2}\hbar^{2}c^{2}%
}{16\epsilon_{1}\epsilon_{2}}\right)  , \label{A31}%
\end{align}
where $d^{2}n_{\pm}=dn_{-}dn_{+}do_{1},d^{2}n_{\gamma}=dn_{\gamma_{1}%
}dn_{\gamma_{2}}do_{-},$ $dn_{\pm}=d\epsilon_{\pm}do_{\pm}\epsilon_{\pm}%
^{2}\beta_{\pm}f_{\pm}$, $dn_{\gamma_{1,2}}=d\epsilon_{1,2}do_{1,2}%
\epsilon_{1,2}^{2}f_{\gamma_{1,2}}$ and the Jacobian is%
\begin{equation}
J_{\mathrm{ca}}=\frac{\epsilon_{+}\beta_{-}}{\left(  \epsilon_{+}+\epsilon
_{-}\right)  \beta_{-}-\left(  \epsilon_{1}\mathbf{b}_{1}+\epsilon
_{2}\mathbf{b}_{2}\right)  \mathbf{\cdot b}_{-}}.
\end{equation}

\subsection{M{\o }ller scattering of electrons and positrons $e_{1}^{\pm}%
e_{2}^{\pm}\rightarrow e_{1}^{\pm\prime}e_{2}^{\pm\prime}$}

\label{moller}

The time evolution of the distribution functions of electrons (or positrons)
is described by%
\begin{equation}
\left(  \frac{\partial f_{i}(\mathbf{p}_{i},t)}{\partial t}\right)
_{e_{1}e_{2}\rightarrow e_{1}^{\prime}e_{2}^{\prime}}=\int d\mathbf{p}%
_{j}d\mathbf{p}_{1}^{\prime}d\mathbf{p}_{2}^{\prime}Vw_{\mathbf{p}_{1}%
^{\prime},\mathbf{p}_{2}^{\prime};\mathbf{p}_{1},\mathbf{p}_{2}}%
[f_{1}(\mathbf{p}_{1}^{\prime},t)f_{2}(\mathbf{p}_{2}^{\prime},t)-f_{1}%
(\mathbf{p}_{1},t)f_{2}(\mathbf{p}_{2},t)]\,, \label{A32}%
\end{equation}
with $i=1,~j=2$, and with $j=1,~i=2$, and where%
\begin{align}
w_{\mathbf{p}_{1}^{\prime},\mathbf{p}_{2}^{\prime};\mathbf{p}_{1}%
,\mathbf{p}_{2}}  &  =\frac{\hbar^{2}c^{6}}{(2\pi)^{2}V}\delta(\epsilon
_{1}+\epsilon_{2}-\epsilon_{1}^{\prime}-\epsilon_{2}^{\prime})\delta
(\mathbf{p}_{1}+\mathbf{p}_{2}-\mathbf{p}_{1}^{\prime}-\mathbf{p}_{2}^{\prime
})\frac{|M_{fi}|^{2}}{16\epsilon_{1}\epsilon_{2}\epsilon_{1}^{\prime}%
\epsilon_{2}^{\prime}},\\
|M_{fi}|^{2}  &  =2^{6}\pi^{2}\alpha^{2}\left\{  \frac{1}{t^{2}}\left[
\frac{s^{2}+u^{2}}{2}+4m^{2}c^{2}(t-m^{2}c^{2})\right]  +\right. \\
&  \left.  +\frac{1}{u^{2}}\left[  \frac{s^{2}+t^{2}}{2}+4m^{2}c^{2}%
(u-m^{2}c^{2})\right]  +\frac{4}{tu}\left(  \frac{s}{2}-m^{2}c^{2}\right)
\left(  \frac{s}{2}-3m^{2}c^{2}\right)  \right\}  , \label{M_fi_e}%
\end{align}
with $s=(\mathfrak{p}_{1}+\mathfrak{p}_{2})^{2}=2(m^{2}c^{2}+\mathfrak{p}%
_{1}\mathfrak{p}_{2})$, $t=(\mathfrak{p}_{1}-\mathfrak{p}_{1}^{\prime}%
)^{2}=2(m^{2}c^{2}-\mathfrak{p}_{1}\mathfrak{p}_{1}^{\prime})$, and
$u=(\mathfrak{p}_{1}-\mathfrak{p}_{2}^{\prime})^{2}=2(m^{2}c^{2}%
-\mathfrak{p}_{1}\mathfrak{p}_{2}^{\prime})$ \cite{1982els..book.....B}.

The energies of final-state particles are given by (\ref{A27}) with new
coefficients $\tilde{A}=(\epsilon_{1}+\epsilon_{2})^{2}-(\epsilon_{1}\beta
_{1}\mathbf{b}_{1}\mathbf{\cdot b}_{1}^{\prime}+\epsilon_{2}\beta
_{2}\mathbf{b}_{2}\mathbf{\cdot b}_{1}^{\prime})^{2}$, $\tilde{B}%
=(\epsilon_{1}+\epsilon_{2})[m^{2}c^{4}+\epsilon_{1}\epsilon_{2}(1-\beta
_{1}\beta_{2}\mathbf{b}_{1}\mathbf{b}_{2})]$, and $\tilde{C}=m^{2}%
c^{4}(\epsilon_{1}\beta_{1}\mathbf{b}_{1}\mathbf{\cdot b}_{1}^{\prime
}+\epsilon_{2}\beta_{2}\mathbf{b}_{2}\mathbf{\cdot b}_{1}^{\prime})^{2}%
+[m^{2}c^{4}+\epsilon_{1}\epsilon_{2}(1-\beta_{1}\beta_{2}\mathbf{b}%
_{1}\mathbf{\cdot b}_{2})]^{2}$. The condition to be checked is%
\begin{equation}
\left[  \epsilon_{1}^{\prime}(\epsilon_{1}+\epsilon_{2})-m^{2}c^{4}%
-(\epsilon_{1}\epsilon_{2})(1-\beta_{1}\beta_{2}\mathbf{b}_{1}\mathbf{\cdot
b}_{2})\right]  \left[  (\epsilon_{1}\beta_{1}\mathbf{b}_{1}+\epsilon_{2}%
\beta_{2}\mathbf{b}_{2})\mathbf{\cdot b}_{1}^{\prime}\right]  \geq0.
\label{A34}%
\end{equation}

Integration of equations (\ref{A32}), similar to the case of Compton
scattering in Section \ref{compton} yields
\begin{align}
\eta_{e,\omega}^{e_{1}e_{2}\rightarrow e_{1}^{\prime}e_{2}^{\prime}}  &
=\frac{1}{\Delta\epsilon_{e,\omega}}\left(  \int_{\epsilon_{1}^{\prime}%
\in\Delta\epsilon_{e,\omega}}d^{2}nJ_{\mathrm{ms}}\frac{\epsilon_{1}^{\prime
2}\beta_{1}^{\prime}|M_{fi}|^{2}\hbar^{2}c^{2}}{16\epsilon_{1}\epsilon
_{2}\epsilon_{2}^{\prime}}+\int_{\epsilon_{2}^{\prime}\in\Delta\epsilon
_{e,\omega}}d^{2}nJ_{\mathrm{ms}}\frac{\epsilon_{1}^{\prime}\beta_{1}^{\prime
}|M_{fi}|^{2}\hbar^{2}c^{2}}{16\epsilon_{1}\epsilon_{2}}\right)  ,\\
(\chi E)_{e,\omega}^{e_{1}e_{2}\rightarrow e_{1}^{\prime}e_{2}^{\prime}}  &
=\frac{1}{\Delta\epsilon_{e,\omega}}\left(  \int_{\epsilon_{1}\in
\Delta\epsilon_{e,\omega}}d^{2}nJ_{\mathrm{ms}}\frac{\epsilon_{1}^{\prime
}\beta_{1}^{\prime}|M_{fi}|^{2}\hbar^{2}c^{2}}{16\epsilon_{2}\epsilon
_{2}^{\prime}}+\int_{\epsilon_{2}\in\Delta\epsilon_{e,\omega}}d^{2}%
nJ_{\mathrm{ms}}\frac{\epsilon_{1}^{\prime}\beta_{1}^{\prime}|M_{fi}|^{2}%
\hbar^{2}c^{2}}{16\epsilon_{1}\epsilon_{2}^{\prime}}\right)  ,
\end{align}
where $d^{2}n=dn_{1}dn_{2}do_{1}^{\prime}$, $dn_{1,2}=d\epsilon_{1,2}%
do_{1,2}\epsilon_{1,2}^{2}\beta_{1,2}f_{_{1,2}}$, and the Jacobian is%
\begin{equation}
J_{\mathrm{ms}}=\frac{\epsilon_{2}^{\prime}\beta_{2}^{\prime}}{(\epsilon
_{1}^{\prime}+\epsilon_{2}^{\prime})\beta_{1}^{\prime}-(\epsilon_{1}\beta
_{1}\mathbf{b}_{1}+\epsilon_{2}\beta_{2}\mathbf{b}_{2})\mathbf{\cdot b}%
_{1}^{\prime}}.
\end{equation}

\subsection{Bhaba scattering of electrons on positrons $e^{-}e^{+}\rightarrow
e^{-\prime}e^{+\prime}$}

The time evolution of the distribution functions of electrons and positrons
due to Bhaba scattering is described by%

\begin{equation}
\left(  \frac{\partial f_{\pm}(\mathbf{p}_{\pm},t)}{\partial t}\right)
_{e^{-}e^{+}\rightarrow e^{-\prime}e^{+\prime}}=\int d\mathbf{p}_{\mp
}d\mathbf{p}_{-}^{\prime}d\mathbf{p}_{+}^{\prime}Vw_{\mathbf{p}_{-}^{\prime
},\mathbf{p}_{+}^{\prime};\mathbf{p}_{-},\mathbf{p}_{+}}[f_{-}(\mathbf{p}%
_{-}^{\prime},t)f_{+}(\mathbf{p}_{+}^{\prime},t)-f_{-}(\mathbf{p}_{-}%
,t)f_{+}(\mathbf{p}_{+},t)], \label{A38}%
\end{equation}
where
\begin{equation}
w_{\mathbf{p}_{-}^{\prime},\mathbf{p}_{+}^{\prime};\mathbf{p}_{-}%
,\mathbf{p}_{+}}=\frac{\hbar^{2}c^{6}}{(2\pi)^{2}V}\delta(\epsilon
_{-}+\epsilon_{+}-\epsilon_{-}^{\prime}-\epsilon_{+}^{\prime})\delta
(\mathbf{p}_{-}+\mathbf{p}_{+}-\mathbf{p}_{-}^{\prime}-\mathbf{p}_{+}^{\prime
})\frac{|M_{fi}|^{2}}{16\epsilon_{-}\epsilon_{+}\epsilon_{-}^{\prime}%
\epsilon_{+}^{\prime}},
\end{equation}
and $|M_{fi}|$ is given by the equation (\ref{M_fi_e}), but the invariants are
$s=(\mathfrak{p}_{-}-\mathfrak{p}_{+}^{\prime})^{2}$, $t=(\mathfrak{p}%
_{+}-\mathfrak{p}_{+}^{\prime})^{2}$ and $u=(\mathfrak{p}_{-}+\mathfrak{p}%
_{+})^{2}$. The final energies $\epsilon_{-}^{\prime}$, $\epsilon_{+}^{\prime
}$ are functions of the outgoing particle directions in a way similar to that
in Section \ref{moller}, see also \cite{1982els..book.....B}.

Integration of equations (\ref{A38}) yields
\begin{align}
\eta_{\pm,\omega}^{e^{-}e^{+}\rightarrow e^{-\prime}e^{+\prime}}  &  =\frac
{1}{\Delta\epsilon_{\pm,\omega}}\left(  \int_{\epsilon_{-}^{\prime}\in
\Delta\epsilon_{e,\omega}}d^{2}n_{\pm}^{\prime}J_{\mathrm{bs}}\frac
{\epsilon_{-}^{\prime2}\beta_{-}^{\prime}|M_{fi}|^{2}\hbar^{2}c^{2}%
}{16\epsilon_{-}\epsilon_{+}\epsilon_{+}^{\prime}}+\int_{\epsilon_{+}^{\prime
}\in\Delta\epsilon_{e,\omega}}d^{2}n_{\pm}^{\prime}J_{\mathrm{bs}}%
\frac{\epsilon_{-}^{\prime}\beta_{-}^{\prime}|M_{fi}|^{2}\hbar^{2}c^{2}%
}{16\epsilon_{-}\epsilon_{+}}\right)  ,\\
(\chi E)_{\pm,\omega}^{e^{-}e^{+}\rightarrow e^{-\prime}e^{+\prime}}  &
=\frac{1}{\Delta\epsilon_{\pm,\omega}}\left(  \int_{\epsilon_{-}\in
\Delta\epsilon_{e,\omega}}d^{2}n_{\pm}^{\prime}J_{\mathrm{bs}}\frac
{\epsilon_{-}^{\prime}\beta_{-}^{\prime}|M_{fi}|^{2}\hbar^{2}c^{2}}%
{16\epsilon_{+}\epsilon_{+}^{\prime}}+\int_{\epsilon_{+}\in\Delta
\epsilon_{e,\omega}}d^{2}n_{\pm}^{\prime}J_{\mathrm{bs}}\frac{\epsilon
_{-}^{\prime}\beta_{-}^{\prime}|M_{fi}|^{2}\hbar^{2}c^{2}}{16\epsilon
_{-}\epsilon_{+}^{\prime}}\right)  ,
\end{align}
where $d^{2}n_{\pm}^{\prime}=dn_{-}dn_{+}do_{-}^{\prime}$, $dn_{\pm}%
=d\epsilon_{\pm}do_{\pm}\epsilon_{\pm}^{2}\beta_{\pm}f_{\pm}$, and the
Jacobian is%
\begin{equation}
J_{\mathrm{bs}}=\frac{\epsilon_{+}^{\prime}\beta_{+}^{\prime}}{(\epsilon
_{-}^{\prime}+\epsilon_{+}^{\prime})\beta_{-}^{\prime}-(\epsilon_{-}\beta
_{-}\mathbf{b}_{-}+\epsilon_{+}\beta_{+}\mathbf{b}_{+})\mathbf{\cdot b}%
_{-}^{\prime}}.
\end{equation}

Analogously to the case of pair creation and annihilation in Section
(\ref{pair}) the energies of final state particles are given by (\ref{A27})
with the coefficients $\breve{A}=(\epsilon_{-}+\epsilon_{+})^{2}-(\epsilon
_{-}\beta_{-}\mathbf{b}_{-}\mathbf{\cdot b}_{-}^{\prime}+\epsilon_{+}\beta
_{+}\mathbf{b}_{+}\mathbf{\cdot b}_{-}^{\prime})^{2}$, $\breve{B}%
=(\epsilon_{-}+\epsilon_{+})\left[  m^{2}c^{4}+\epsilon_{-}\epsilon
_{+}(1-\beta_{-}\beta_{+}\mathbf{b}_{-}\mathbf{\cdot b}_{+})\right]  $,
$\breve{C}=\left[  m^{2}c^{4}+\epsilon_{-}\epsilon_{+}(1-\beta_{-}\beta
_{+}\mathbf{b}_{-}\mathbf{\cdot b}_{+})\right]  ^{2}+m^{2}c^{4}\left[
\epsilon_{-}\beta_{-}\mathbf{b}_{-}\mathbf{\cdot b}_{-}^{\prime}+\epsilon
_{+}\beta_{+}\mathbf{b}_{+}\mathbf{\cdot b}_{-}^{\prime}\right]  ^{2}$. In
order to select the correct root one has to check the condition (\ref{A34})
changing the subscripts $1\rightarrow-$, $2\rightarrow+$.

\section{Binary reactions with protons}

\label{2body_p}

\subsection{Compton scattering on protons $\gamma p\rightarrow\gamma^{\prime
}p^{\prime}$}

The rate for this process $t_{\gamma p}^{-1}$, compared to the rate of Compton
scattering of electrons $t_{\gamma e}^{-1}$ is much longer,%

\begin{equation}
t_{\gamma p}^{-1}=\frac{n_{p}}{n_{\pm}}\left(  \frac{\epsilon_{\pm}}{Mc^{2}%
}\right)  ^{2}t_{\gamma e}^{-1}\qquad\epsilon\geq mc^{2}.
\end{equation}
Moreover, it is longer than any timescale for binary and triple reactions
considered in this paper and thus we exclude this reaction from the computations.

\subsection{Electron-proton and positron-proton scattering $e_{\pm
}p\rightarrow e_{\pm}^{\prime}p^{\prime}$}

The time evolution of the distribution functions of electrons due to
$ep\rightarrow e^{\prime}p^{\prime}$ is described by
\begin{align}
\left(  \frac{\partial f_{\pm}(\mathbf{p},t)}{\partial t}\right)
_{ep\rightarrow e^{\prime}p^{\prime}}  &  =\int d\mathbf{q}d\mathbf{p}%
^{\prime}d\mathbf{q}^{\prime}Vw_{\mathbf{p}^{\prime},\mathbf{q}^{\prime
};\mathbf{p},\mathbf{q}}[f_{\pm}(\mathbf{p}^{\prime},t)f_{p}(\mathbf{q}%
^{\prime},t)-f_{\pm}(\mathbf{p},t)f_{p}(\mathbf{q},t)],\\
\left(  \frac{\partial f_{p}(\mathbf{q},t)}{\partial t}\right)
_{ep\rightarrow e^{\prime}p^{\prime}}  &  =\int d\mathbf{p}d\mathbf{p}%
^{\prime}d\mathbf{q}^{\prime}Vw_{\mathbf{p}^{\prime},\mathbf{q}^{\prime
};\mathbf{p},\mathbf{q}}[f_{\pm}(\mathbf{p}^{\prime},t)f_{p}(\mathbf{q}%
^{\prime},t)-f_{\pm}(\mathbf{p},t)f_{p}(\mathbf{q},t)],
\end{align}
where
\begin{align}
w_{\mathbf{p}^{\prime},\mathbf{q}^{\prime};\mathbf{p},\mathbf{q}}  &
=\frac{\hbar^{2}c^{6}}{(2\pi)^{2}V}\delta(\epsilon_{e}+\epsilon_{p}%
-\epsilon_{e}^{\prime}-\epsilon_{p}^{\prime})\delta(\mathbf{p}+\mathbf{q}%
-\mathbf{p}^{\prime}-\mathbf{q}^{\prime})\frac{|M_{fi}|^{2}}{16\epsilon
_{e}\epsilon_{p}\epsilon_{e}^{\prime}\epsilon_{p}^{\prime}},\\
|M_{fi}|^{2}  &  =2^{6}\pi^{2}\alpha^{2}\frac{1}{t^{2}}\left\{  \frac{1}%
{2}(s^{2}+u^{2})+(m^{2}c^{2}+M^{2}c^{2})(2t-m^{2}c^{2}-M^{2}c^{2})\right\}  ,
\label{M_fi_ep}%
\end{align}
the invariants are $s=(\mathfrak{p}+\mathfrak{q})^{2}=m^{2}c^{2}+M^{2}%
c^{2}+2\mathfrak{p}\cdot\mathfrak{q}$, $t=(\mathfrak{p}-\mathfrak{p}^{\prime
})^{2}=2(m^{2}c^{2}-\mathfrak{p}\cdot\mathfrak{p}^{\prime})=2(M^{2}%
c^{2}-\mathfrak{q}\cdot\mathfrak{q}^{\prime})$ and $u=(\mathfrak{p}%
-\mathfrak{q}^{\prime})^{2}=m^{2}c^{2}+M^{2}c^{2}-2\mathfrak{p}\cdot
\mathfrak{q}^{\prime}$, $s+t+u=2(m^{2}c^{2}+M^{2}c^{2})$. The energies of
particles after interaction are given by (\ref{A27}) with $\bar{A}%
=(\epsilon_{\pm}+\epsilon_{p})^{2}-\left[  (\epsilon_{\pm}\beta_{\pm
}\mathbf{b}_{\pm}+\epsilon_{p}\beta_{p}\mathbf{b}_{p})\mathbf{\cdot b}_{\pm
}^{\prime}\right]  ^{2}$, $\bar{B}=(\epsilon_{\pm}+\epsilon_{p})[m^{2}%
c^{4}+\epsilon_{\pm}\epsilon_{p}(1-\beta_{\pm}\beta_{p}\mathbf{b}_{\pm
}\mathbf{\cdot b}_{p})]$, $\bar{C}=m^{2}c^{4}\left\{  (\epsilon_{\pm}%
\beta_{\pm}\mathbf{b}_{\pm}\mathbf{\cdot b}_{\pm}^{\prime}+\epsilon_{p}%
\beta_{p}\mathbf{b}_{p}\mathbf{\cdot b}_{\pm}^{\prime})^{2}+[m^{2}%
c^{4}+\epsilon_{\pm}\epsilon_{p}(1-\beta_{\pm}\beta_{p}\mathbf{b}_{\pm
}\mathbf{\cdot b}_{p})]\right\}  ^{2}$. The correct root is selected by the
condition (\ref{A34}) with the substitution $1\rightarrow\pm$, $2\rightarrow
p$.

Absorption and emission coefficients for this reaction are%
\begin{align}
(\chi E)_{\pm,\omega}^{ep}  &  =\frac{1}{\Delta\epsilon_{\pm,\omega}}%
\int_{\epsilon_{\pm}\in\Delta\epsilon_{\pm,\omega}}dn_{\pm}dn_{p}do_{\pm
}^{\prime}J_{\mathrm{ep}}\frac{\epsilon_{\pm}^{\prime2}\beta_{\pm}^{\prime
}\epsilon_{\pm}|M_{fi}|^{2}\hbar^{2}c^{2}}{16\epsilon_{\pm}\epsilon
_{p}\epsilon_{\pm}^{\prime}\epsilon_{p}^{\prime}},\\
(\chi E)_{p,\omega}^{ep}  &  =\frac{1}{\Delta\epsilon_{p,\omega}}%
\int_{\epsilon_{p}\in\Delta\epsilon_{p,\omega}}dn_{\pm}dn_{p}do_{\pm}^{\prime
}J_{\mathrm{ep}}\frac{\epsilon_{\pm}^{\prime2}\beta_{\pm}^{\prime}\epsilon
_{p}|M_{fi}|^{2}\hbar^{2}c^{2}}{16\epsilon_{\pm}\epsilon_{p}\epsilon_{\pm
}^{\prime}\epsilon_{p}^{\prime}},\\
\eta_{\pm,\omega}^{ep}  &  =\frac{1}{\Delta\epsilon_{\pm,\omega}}%
\int_{\epsilon_{\pm}^{\prime}\in\Delta\epsilon_{\pm,\omega}}dn_{\pm}%
dn_{p}do_{\pm}^{\prime}J_{\mathrm{ep}}\frac{\epsilon_{\pm}^{\prime2}\beta
_{\pm}^{\prime}\epsilon_{\pm}^{\prime}|M_{fi}|^{2}\hbar^{2}c^{2}}%
{16\epsilon_{\pm}\epsilon_{p}\epsilon_{\pm}^{\prime}\epsilon_{p}^{\prime}},\\
\eta_{p,\omega}^{ep}  &  =\frac{1}{\Delta\epsilon_{p,\omega}}\int
_{\epsilon_{p}^{\prime}\in\Delta\epsilon_{p,\omega}}dn_{\pm}dn_{p}do_{\pm
}^{\prime}J_{\mathrm{ep}}\frac{\epsilon_{\pm}^{\prime2}\beta_{\pm}^{\prime
}\epsilon_{p}^{\prime}|M_{fi}|^{2}\hbar^{2}c^{2}}{16\epsilon_{\pm}\epsilon
_{p}\epsilon_{\pm}^{\prime}\epsilon_{p}^{\prime}},
\end{align}
where $dn_{i}=d\epsilon_{i}do_{i}\epsilon_{i}^{2}\beta_{i}f_{i}$, $i=\pm,p$,
and the Jacobian is
\begin{equation}
J_{\mathrm{ep}}=\frac{\epsilon_{p}^{\prime}\beta_{p}^{\prime}}{(\epsilon_{\pm
}^{\prime}+\epsilon_{p}^{\prime})\beta_{\pm}^{\prime}-(\epsilon_{p}\beta
_{p}\mathbf{b}_{p}+\epsilon_{\pm}\beta_{\pm}\mathbf{b}_{\pm})\mathbf{\cdot
b}_{\pm}^{\prime}}.
\end{equation}

The rate for proton-electron (proton-positron) scattering is%

\begin{equation}
t_{ep}^{-1}\approx\frac{\epsilon}{Mc^{2}}t_{ee}^{-1},\qquad\epsilon_{\pm}%
\ll\epsilon_{p}. \label{epR}%
\end{equation}

\subsection{Proton-proton scattering $p_{1}p_{2}\rightarrow p_{1}^{\prime
}p_{2}^{\prime}$}

This reaction is similar to $e_{1}e_{2}\rightarrow e_{1}^{\prime}e_{2}%
^{\prime}$, described in Section \ref{moller}. The time evolution of the
distribution functions of electrons is described by
\begin{equation}
\left(  \frac{\partial f_{i}(\mathbf{p}_{i},t)}{\partial t}\right)
_{p_{1}p_{2}\rightarrow p_{1}^{\prime}p_{2}^{\prime}}=\int d\mathbf{q}%
_{j}d\mathbf{q}_{1}^{\prime}d\mathbf{q}_{2}^{\prime}Vw_{\mathbf{q}_{1}%
^{\prime},\mathbf{q}_{2}^{\prime};\mathbf{q}_{1},\mathbf{q}_{2}}%
[f_{1}(\mathbf{q}_{1}^{\prime},t)f_{2}(\mathbf{q}_{2}^{\prime},t)-f_{1}%
(\mathbf{q}_{1},t)f_{2}(\mathbf{q}_{2},t)],
\end{equation}
with $j=3-i$, and where
\begin{align}
w_{\mathbf{q}_{1}^{\prime},\mathbf{q}_{2}^{\prime};\mathbf{q}_{1}%
,\mathbf{q}_{2}}  &  =\frac{\hbar^{2}c^{6}}{(2\pi)^{2}V}\delta(\epsilon
_{1}+\epsilon_{2}-\epsilon_{1}^{\prime}-\epsilon_{2}^{\prime})\delta
(\mathbf{q}_{1}+\mathbf{q}_{2}-\mathbf{q}_{1}^{\prime}-\mathbf{q}_{2}^{\prime
})\frac{|M_{fi}|^{2}}{16\epsilon_{1}\epsilon_{2}\epsilon_{1}^{\prime}%
\epsilon_{2}^{\prime}},\\
|M_{fi}|^{2}  &  =2^{6}\pi^{2}\alpha^{2}\left\{  \frac{1}{t^{2}}\left[
\frac{s^{2}+u^{2}}{2}+4M^{2}c^{2}(t-M^{2}c^{2})\right]  +\right. \nonumber\\
&  \left.  \frac{1}{u^{2}}\left[  \frac{s^{2}+t^{2}}{2}+4M^{2}c^{2}%
(u-M^{2}c^{2})\right]  +\frac{4}{tu}\left(  \frac{s}{2}-M^{2}c^{2}\right)
\left(  \frac{s}{2}-3M^{2}c^{2}\right)  \right\}  , \label{M_fi_p}%
\end{align}
and the invariants are $s=(\mathfrak{q}_{1}+\mathfrak{q}_{2})^{2}=2(M^{2}%
c^{2}+\mathfrak{q}_{1}\cdot\mathfrak{q}_{2})$, $t=(\mathfrak{q}_{1}%
-\mathfrak{q}_{1}^{\prime})^{2}=2(M^{2}c^{2}-\mathfrak{q}_{1}\cdot
\mathfrak{q}_{1}^{\prime})$, and $u=(\mathfrak{q}_{1}-\mathfrak{q}_{2}%
^{\prime})^{2}=2(M^{2}c^{2}-\mathfrak{q}_{1}\mathfrak{q}_{2}^{\prime})$.

For the rate we have%

\begin{equation}
t_{pp}^{-1}\approx\sqrt{\frac{m}{M}}\frac{n_{p}}{n_{\pm}}t_{ee}^{-1},\qquad
v_{p}\approx\sqrt{\frac{m}{M}}v_{\pm},\qquad v_{\pm}\approx c. \label{ppR}%
\end{equation}

\section{Three-body processes}

\label{3body}

We adopt emission coefficients for triple interactions from
\cite{1984MNRAS.209..175S}.

Bremsstrahlung%
\begin{equation}
\eta_{\gamma}^{e^{\mp}e^{\mp}\rightarrow e^{\mp}e^{\mp}\gamma}=(n_{+}%
^{2}+n_{-}^{2})\frac{16}{3}\frac{\alpha c}{\varepsilon}\left(  \frac{e^{2}%
}{mc^{2}}\right)  ^{2}\ln\left[  4\xi(11.2+10.4\theta^{2})\frac{\theta
}{\varepsilon}\right]  \frac{\frac{3}{5}\sqrt{2}\theta+2\theta^{2}}%
{\exp(1/\theta)K_{2}(1/\theta)},
\end{equation}%
\begin{equation}
\eta_{\gamma}^{e^{-}e^{+}\rightarrow e^{-}e^{+}\gamma}=n_{+}n_{-}\frac{16}%
{3}\frac{2\alpha c}{\varepsilon}\left(  \frac{e^{2}}{mc^{2}}\right)  ^{2}%
\ln\left[  4\xi(1+10.4\theta^{2})\frac{\theta}{\varepsilon}\right]
\frac{\sqrt{2}+2\theta+2\theta^{2}}{\exp(1/\theta)K_{2}(1/\theta)},
\end{equation}%
\begin{equation}
\eta_{\gamma}^{pe^{\pm}\rightarrow p^{\prime}e^{\pm\prime}\gamma}=(n_{+}%
+n_{-})n_{p}\frac{16}{3}\frac{\alpha c}{\varepsilon}\left(  \frac{e^{2}%
}{mc^{2}}\right)  ^{2}\ln\left[  4\xi(1+3.42\theta)\frac{\theta}{\varepsilon
}\right]  \frac{1+2\theta+2\theta^{2}}{\exp(1/\theta)K_{2}(1/\theta)},
\end{equation}
where $\xi=e^{-0.5772}$, and $K_{2}(1/\theta)$ is the modified Bessel function
of the second kind of order 2.

Double Compton scattering%
\begin{equation}
\eta_{\gamma}^{e^{\pm}\gamma\rightarrow e^{\pm\prime}\gamma^{\prime}%
\gamma^{\prime\prime}}=(n_{+}+n_{-})n_{\gamma}\frac{128}{3}\frac{\alpha
c}{\varepsilon}\left(  \frac{e^{2}}{mc^{2}}\right)  ^{2}\frac{\theta^{2}%
}{1+13.91\theta+11.05\theta^{2}+19.92\theta^{3}},
\end{equation}

Three photon annihilation%

\begin{equation}
\eta_{\gamma}^{e^{\pm}e^{\mp}\rightarrow\gamma\gamma^{\prime}\gamma
^{\prime\prime}}=n_{+}n_{-}\alpha c\left(  \frac{e^{2}}{mc^{2}}\right)
^{2}\frac{1}{\varepsilon}\frac{\frac{4}{\theta}\left(  2\ln^{2}2\xi
\theta+\frac{\pi^{2}}{6}-\frac{1}{2}\right)  }{4\theta+\frac{1}{\theta^{2}%
}\left(  2\ln^{2}2\xi\theta+\frac{\pi^{2}}{6}-\frac{1}{2}\right)  },
\end{equation}
where we have joined two limiting approximations given by
\cite{1984MNRAS.209..175S}.

Radiative pair production%

\begin{equation}
\eta_{e}^{\gamma\gamma^{\prime}\rightarrow\gamma^{\prime\prime}e^{\pm}e^{\mp}%
}=\eta_{\gamma}^{e^{\pm}e^{\mp}\rightarrow\gamma\gamma^{\prime}\gamma
^{\prime\prime}}\frac{n_{\gamma}^{2}}{n_{+}n_{-}}\left[  \frac{K_{2}%
(1/\theta)}{2\theta^{2}}\right]  ^{2}.
\end{equation}

Electron-photon pair production%
\begin{equation}
\eta_{\gamma}^{e_{1}^{\pm}\gamma\rightarrow e_{1}^{\pm\prime}e^{\pm}e^{\mp}%
}=\left\{
\begin{array}
[c]{cc}%
(n_{+}+n_{-})n_{\gamma}\alpha c\left(  \frac{e^{2}}{mc^{2}}\right)  ^{2}%
\exp\left(  -\frac{2}{\theta}\right)  16.1\theta^{0.541}, & \theta\leq2,\\
(n_{+}+n_{-})n_{\gamma}\alpha c\left(  \frac{e^{2}}{mc^{2}}\right)
^{2}\left(  \frac{56}{9}\ln2\xi\theta-\frac{8}{27}\right)  \frac
{1}{1+0.5/\theta}, & \theta>2.
\end{array}
\right.
\end{equation}
Proton-photon pair production%
\begin{equation}
\eta_{\gamma}^{p\gamma\rightarrow p^{\prime}e^{\pm}e^{\mp}}=\left\{
\begin{array}
[c]{cc}%
n_{p}n_{\gamma}\alpha c\left(  \frac{e^{2}}{mc^{2}}\right)  ^{2}\exp\left(
-\frac{2}{\theta}\right)  \frac{1}{1+0.9\theta}, & \theta\leq1.25277,\\
n_{p}n_{\gamma}\alpha c\left(  \frac{e^{2}}{mc^{2}}\right)  ^{2}\left[
\frac{28}{9}\left(  \ln2\xi\theta+1.7\right)  -\frac{92}{27}\right]  , &
\theta>1.25277.
\end{array}
\right.  .
\end{equation}

We use the absorption coefficient for three-body processes written as
\begin{equation}
\chi_{\gamma}^{\mathrm{3p}}=\eta_{\gamma}^{\mathrm{3p}}/E_{\gamma
}^{\mathrm{eq}}\,,
\end{equation}
where $\eta_{\gamma}^{\mathrm{3p}}$ is the sum of the emission coefficients of
photons in the three particle processes, $E_{\gamma}^{\mathrm{eq}}%
=2\pi\epsilon^{3}f_{\gamma}^{\mathrm{eq}}/c^{3}$, where $f_{\gamma
}^{\mathrm{eq}}$ is given by (\ref{dk}).

From equation (\ref{EBoltzmannEq}), the law of energy conservation in the
three-body processes is
\begin{equation}
\int{\sum_{i}(\eta_{i}^{\mathrm{3p}}-\chi_{i}^{\mathrm{3p}}E_{i})d\mu
d\epsilon}=0\,.
\end{equation}
For exact conservation of energy in these processes we introduce the following
coefficients of emission and absorption for electrons:
\begin{equation}
\chi_{e}^{\mathrm{3p}}=\frac{\int(\eta_{\gamma}^{\mathrm{3p}}-\chi_{\gamma
}^{\mathrm{3p}}E_{\gamma})d\epsilon d\mu}{\int E_{e}d\epsilon d\mu},\qquad
\eta_{e}^{\mathrm{3p}}=0,\qquad\int(\eta_{\gamma}^{\mathrm{3p}}-\chi_{\gamma
}^{\mathrm{3p}}E_{\gamma})d\epsilon d\mu>0\,,
\end{equation}
or
\begin{equation}
\frac{\eta_{e}^{\mathrm{3p}}}{E_{e}}=-\frac{\int(\eta_{\gamma}^{\mathrm{3p}%
}-\chi_{\gamma}^{\mathrm{3p}}E_{\gamma})d\epsilon d\mu}{\int E_{e}d\epsilon
d\mu},\qquad\chi_{e}^{\mathrm{3p}}=0,\qquad\int(\eta_{\gamma}^{\mathrm{3p}%
}-\chi_{\gamma}^{\mathrm{3p}}E_{\gamma})d\epsilon d\mu<0\,.
\end{equation}
%

\end{widetext}%

\section{Cutoff in the Coulomb scattering}

\label{cutoff}

Denote quantities in the center of mass (CM) frame with index $0$, and with
prime after interaction. Suppose we have two particles with masses $m_{1}$ and
$m_{2}$. The change of the angle of the first particle in CM system is
\begin{equation}
\theta_{10}=\arccos(\mathbf{b}_{10}\mathbf{\cdot b}_{10}^{\prime}),
\label{angle}%
\end{equation}
the numerical grid size is $\Delta\theta_{\mathrm{g}}$, the minimal angle at
the scattering is $\theta_{\mathrm{min}}$.

By definition in the in CM frame%
\begin{equation}
\mathbf{p}_{10}+\mathbf{p}_{20}=0,
\end{equation}
where%
\begin{equation}
\mathbf{p}_{i0}=\mathbf{p}_{i}+\left[  (\Gamma-1)(\mathbf{N}\mathbf{p}%
_{i})-\Gamma\frac{V}{c}\frac{\epsilon_{i}}{c}\right]  \mathbf{N},\quad i=1,2,
\end{equation}
and%
\begin{equation}
\epsilon_{i}=\Gamma(\epsilon_{i0}+\mathbf{V}\mathbf{p}_{i0}).
\end{equation}
Then for the velocity of the CM\ frame we have
\begin{equation}
\frac{\mathbf{V}}{c}=c\frac{\mathbf{p}_{1}+\mathbf{p}_{2}}{\epsilon
_{1}+\epsilon_{2}},\quad\mathbf{N}=\frac{\mathbf{V}}{V},\quad\Gamma=\frac
{1}{\sqrt{1-\left(  \frac{V}{c}\right)  ^{2}}}.
\end{equation}
By definition
\begin{equation}
\mathbf{b}_{10}=\mathbf{b}_{20},\quad\mathbf{b}_{10}^{\prime}=\mathbf{b}%
_{20}^{\prime},
\end{equation}
and then%
\begin{gather}
\left\vert \mathbf{p}_{10}\right\vert =\left\vert \mathbf{p}_{20}\right\vert
=p_{0}\equiv\nonumber\\
\equiv\frac{1}{c}\sqrt{\epsilon_{10}^{2}-m_{1}^{2}c^{4}}=\frac{1}{c}%
\sqrt{\epsilon_{20}^{2}-m_{2}^{2}c^{4}},
\end{gather}
where%
\begin{align}
\epsilon_{10}  &  =\frac{(\epsilon_{1}+\epsilon_{2})^{2}-\Gamma^{2}(m_{2}%
^{2}-m_{1}^{2})c^{4}}{2(\epsilon_{1}+\epsilon_{2})\Gamma},\\
\epsilon_{20}  &  =\frac{(\epsilon_{1}+\epsilon_{2})^{2}+\Gamma^{2}(m_{2}%
^{2}-m_{1}^{2})c^{4}}{2(\epsilon_{1}+\epsilon_{2})\Gamma}.
\end{align}
Haug \cite{1988A&A...191..181H} gives the minimal scattering angle in the
center of mass system%
\begin{equation}
\theta_{\mathrm{min}}=\frac{2\hbar}{\mathcal{M}cD}\frac{\gamma_{r}}%
{(\gamma_{r}+1)\sqrt{2(\gamma_{r}-1)}},
\end{equation}
where $\mathcal{M}$, as above, is the reduced mass, the maximum impact
parameter (neglecting the effect of protons) is
\begin{equation}
D=\frac{c^{2}}{\omega}\frac{p_{0}}{\epsilon_{10}},
\end{equation}
and the invariant Lorentz factor of relative motion (e.g.
\cite{1988A&A...191..181H}) is%
\begin{equation}
\gamma_{r}=\frac{1}{\sqrt{1-\left(  \frac{v_{r}}{c}\right)  ^{2}}}%
=\frac{\epsilon_{1}\epsilon_{2}-\mathbf{p}_{1}\mathbf{p}_{2}c^{2}}{m_{1}%
m_{2}c^{4}}. \label{gammarel}%
\end{equation}

In the CM frame we finally obtain%
\[
t_{\mathrm{min}}=2\left[  \left(  mc\right)  ^{2}-\left(  \frac{\epsilon_{10}%
}{c}\right)  ^{2}\left(  1-\beta_{10}^{2}\cos\theta_{\mathrm{min}}\right)
\right]
\]

Since it is invariant, we then replace $t$ in the denominator of $|M_{fi}%
|^{2}$ in (\ref{M_fi_e}) by the value $t\sqrt{1+t_{\mathrm{min}}^{2}/t^{2}}$
to implement the cutoff scheme. Also at the scattering of equivalent particles
we remove the case of exchange of particles as well as scattering on small
angles, in other words we change $u$ in the denominator of $|M_{fi}|^{2}$ in
(\ref{M_fi_e}),(\ref{M_fi_ep}) and (\ref{M_fi_p}) by the value $u\sqrt
{1+t_{\mathrm{min}}^{2}/u^{2}}$.

\section{Mass scaling for the proton-electron/positron reaction}

\label{massscaling}

Since proton mass is larger than electron mass-energy $M\gg m,\epsilon$ then
for the CM frame%
\begin{gather}
\mathbf{V\thickapprox}\frac{\mathbf{p}_{1}+\mathbf{p}_{2}}{M},\quad
\Gamma\thickapprox1,\quad J_{1}\thickapprox1,\\
\epsilon_{1}^{\prime}-\epsilon_{1}\thickapprox\mathbf{V}\left(  \mathbf{e}%
_{01}^{\prime}-\mathbf{e}_{01}\right)  p_{0}\propto\frac{1}{M},
\end{gather}
and also%

\begin{gather}
\frac{s^{2}}{c^{4}}\thickapprox M^{4}+4mM^{3}+6m^{2}M^{2},\\
\frac{u^{2}}{c^{4}}\thickapprox M^{4}-4mM^{3}+6m^{2}M^{2},\\
\left\vert M_{fi}\right\vert ^{2}\propto\frac{1}{t^{2}}\left(  6m^{2}%
-2t\right)  M^{2},
\end{gather}
while%

\begin{align}
t  &  =\frac{-2m^{2}\beta_{e0}^{2}\left(  1-\mathbf{e}_{e0}\mathbf{e}%
_{e0}^{\prime}\right)  }{1-\beta_{e0}^{2}}=\\
&  =\frac{-2m^{2}\beta_{e}^{2}\left(  1-\mathbf{e}_{e}\mathbf{e}_{e}^{\prime
}\right)  }{1-\beta_{e}^{2}}\left[  1+O\left(  M^{-1}\right)  \right]
\end{align}
for small angles.

This leads to the following scaling for the reaction rate%

\begin{equation}
\eta_{e\omega}^{ep}-\left(  \chi E\right)  _{e\omega}^{ep}\propto\int
\frac{\left(  \epsilon_{e}^{\prime}-\epsilon_{e}\right)  \left\vert
M_{fi}\right\vert ^{2}}{\epsilon_{e}\epsilon_{p}\epsilon_{e}^{\prime}%
\epsilon_{p}^{\prime}}\propto\frac{1}{M.}%
\end{equation}

We can therefore calculate $\eta_{e\omega}^{ep_{0}}$, $\left(  \chi E\right)
_{e\omega}^{ep_{0}}$ for a pseudo-particle with mass $M_{0}\gg m$, $\epsilon$
instead of $M$ and obtain%

\begin{align}
\eta_{e\omega}^{ep}  &  \thickapprox\frac{M_{0}}{M}\eta_{e\omega}^{ep_{0}},\\
\left(  \chi E\right)  _{e\omega}^{ep}  &  \thickapprox\frac{M_{0}}{M}\left(
\chi E\right)  _{e\omega}^{ep_{0}}.
\end{align}

For such purpose we selected the mass of this pseudo-particle as $M_{0}=20m$.

\section{The definition of matrix elements}

\label{ME}

Following \cite{1982els..book.....B}\ define the scattering matrix, being
composed of real and imaginary parts%
\begin{equation}
S_{fi}=\delta_{fi}+i\left(  2\pi\hbar\right)  ^{4}\delta^{(4)}\left(
\mathfrak{p}_{f}-\mathfrak{p}_{i}\right)  T_{fi},
\end{equation}
where $\delta_{fi}$ is the unity matrix, $\delta^{(4)}$ stands for the
four-momentum conservation and the elements of $T_{fi}$ are scattering amplitudes.

The transition probability of a given process per unit time is then%
\begin{equation}
w_{fi}=c\left(  2\pi\hbar\right)  ^{4}\delta^{(4)}\left(  \mathfrak{p}%
_{f}-\mathfrak{p}_{i}\right)  \left\vert T_{fi}\right\vert ^{2}V,
\end{equation}
where $V$ is the normalization volume.

For a process involving $a$\ outgoing particles and $b$ incoming particles the
differential probability per unit time is defined as%
\begin{align}
dw  &  =c(2\pi\hbar)^{4}\delta^{(4)}\left(  \mathfrak{p}_{f}-\mathfrak{p}%
_{i}\right)  \left\vert M_{fi}\right\vert ^{2}V\times\label{dp}\\
&  \times\left[  \prod\limits_{b}\frac{\hbar c}{2\epsilon_{b}V}\right]
\left[  \prod_{a}\frac{d\mathbf{p}_{a}^{\prime}}{(2\pi\hbar)^{3}}\frac{\hbar
c}{2\epsilon_{a}^{\prime}}\right]  ,\nonumber
\end{align}
where $\mathbf{p}_{a}^{\prime}$ and $\epsilon_{a}^{\prime}$ are respectively
momenta and energies of outgoing particles, $\epsilon_{b}$ are energies of
particles before interaction, $M_{fi}$ are the corresponding matrix elements,
$\delta^{(4)}$ stands for energy-momentum conservation, $V$ is the
normalization volume. The matrix elements are related to the scattering
amplitudes by%
\begin{equation}
M_{fi}=\left[  \prod\limits_{b}\frac{\hbar c}{2\epsilon_{b}V}\right]  \left[
\prod_{a}\frac{\hbar c}{2\epsilon_{a}^{\prime}V}\right]  T_{fi}.
\end{equation}

For a binary process with 2 incoming and 2 outgoing particles it is convenient
to introduce the differential cross-section. In fact, the differential
probability for incoming particles with four momenta $\mathfrak{p}_{1}$ and
$\mathfrak{p}_{2}$, energies $\epsilon_{1}$\ and $\epsilon_{2}$\ and masses
$m_{1}$\ and $m_{2}$\ respectively, is just the product of the differential
cross-section and the flux density%
\begin{equation}
dw=jd\sigma,
\end{equation}
where
\begin{align}
j  &  =\frac{cI}{\epsilon_{1}\epsilon_{2}V},\\
I  &  =c\sqrt{\mathfrak{p}_{1}\mathfrak{p}_{2}-m_{1}m_{2}c^{2}}.
\end{align}
In the CM\ reference frame the relation between the cross section and
$\left\vert M_{fi}\right\vert ^{2}$ acquires simplest form if cross-section is
independent on the azimuth of $\mathbf{p}_{1}^{\prime}$\ relative to
$\mathbf{p}_{1}$ then%
\begin{align}
d\sigma &  =\frac{\hbar^{2}c^{4}}{64\pi}\left\vert M_{fi}\right\vert ^{2}%
\frac{dt}{I},\\
t  &  =\left(  \mathfrak{p}_{1}-\mathfrak{p}_{2}\right)  ^{2},\\
dt  &  =2\left\vert \mathbf{p}_{1}\right\vert \left\vert \mathbf{p}%
_{1}^{\prime}\right\vert d\cos\vartheta,
\end{align}
where $\vartheta$ is the angle between $\mathbf{p}_{1}$ and $\mathbf{p}%
_{1}^{\prime}$.

\bibliographystyle{aip}
\bibliography{pair}

\end{document}